\documentclass[a4paper,11pt]{article}
\pdfoutput=1 

\usepackage{jcappub} 

\usepackage[T1]{fontenc} 
\usepackage{natbib} 
\usepackage[sort&compress,numbers]{natbib}
\usepackage{multirow}

\newcommand{\Lim}[1]{\raisebox{0.5ex}{\scalebox{0.8}{$\displaystyle \lim_{#1}\;$}}}

\title{\boldmath Biases on cosmological parameter estimators from galaxy cluster number counts}

\author[a,b]{M. Penna-Lima}
\author[b]{M. Makler,}
\author[a]{C. A. Wuensche}

\affiliation[a]{Instituto Nacional de Pesquisas Espaciais, Divis\~ao de Astrof\'isica, \\ Av. dos Astronautas 1758, 12227-010, S\~ao Jos\'e dos Campos -- SP, Brazil}
\affiliation[b]{Centro Brasileiro de Pesquisas F\'{\i}sicas,\\ Rua Dr.\ Xavier Sigaud 150,
22290-180, Rio de Janeiro -- RJ, Brazil}

\emailAdd{mariana.lima@inpe.br}
\emailAdd{martin@cbpf.br}
\emailAdd{ca.wuensche@inpe.br}

\abstract{Sunyaev-Zel'dovich (SZ) surveys are promising probes of cosmology - in particular for Dark Energy (DE) -, given their ability to find distant clusters and provide estimates for their mass.  However, current SZ catalogs contain tens to hundreds of objects and maximum likelihood estimators may present biases for such sample sizes. In this work we study estimators from cluster abundance for some cosmological parameters, in particular the DE equation of state parameter $w_0$, the amplitude of density fluctuations  $\sigma_8$, and the Dark Matter density parameter $\Omega_c$. We begin by deriving an unbinned likelihood for cluster number counts, showing that it is equivalent to the one commonly used in the literature. We use the Monte Carlo approach to determine the presence of bias using this likelihood and study its behavior with both the area and depth of the survey, and the number of cosmological parameters fitted. Our fiducial models are based on the South Pole Telescope (SPT) SZ survey. Assuming perfect knowledge of mass and redshift some estimators have non-negligible biases. For example, the bias of $\sigma_8$ corresponds to about $40\%$ of its statistical error bar when fitted together with $\Omega_c$  and $w_0$. Including a SZ mass-observable relation decreases the relevance of the bias, for the typical sizes of current SZ surveys. Considering a joint likelihood for cluster abundance and the so-called ``distance priors'', we obtain that the biases are negligible compared to the statistical errors. However, we show that the biases from SZ estimators do not go away with increasing sample sizes and they may become the dominant source of error for an all sky survey at the SPT sensitivity. Finally, we compute the confidence regions for the cosmological parameters using Fisher matrix and profile likelihood approaches, showing that they are compatible with the Monte Carlo ones. The results of this work validate the use of the current maximum likelihood methods for present SZ surveys, but highlight the need for further studies for upcoming experiments. To perform the analyses of this work, we developed fast, accurate, and adaptable codes for cluster counts in the framework of the Numerical Cosmology Library.}


\arxivnumber{1312.4430}

\begin{document}
\maketitle
\flushbottom

\section{Introduction}
\label{sec:introduction}

The abundance of galaxy clusters and their spatial distribution as functions of redshift and mass provide strong constraints on the matter density parameter $\Omega_m$ and the amplitude of the matter power spectrum $\sigma_8$~\cite{Holder2001, Haiman2001, Borgani2001, Huterer2002, Vikhlinin2009a}. For high redshift surveys ($z > 1.0$), clusters are also powerful tools to study Dark Energy (DE) since they depend on the linear growth of density perturbations and the comoving volume~\cite{Haiman2001, Mantz2010} (for a review see~\cite{Allen2011} and references therein). The coming years are promising for cluster cosmology in view of surveys which will provide deep and large catalogs such as the Dark Energy Survey (DES)~\cite{Abbott2005} and the forthcoming Large Synoptic Survey Telescope (LSST)~\cite{LSST2012} optical surveys. The ongoing Sunyaev-Zel'dovich (SZ) surveys carried by the South Pole Telescope (SPT)~\cite{Vanderlinde2010, Benson2013, Reichardt2013}, the Atacama Cosmology Telescope (ACT)~\cite{Sehgal2011} and the Planck probe~\cite{Douspis2013} have already yielded catalogs with massive clusters and redshifts up to $1.5$.

The number density of Dark Matter (DM) halos as a function of redshift and mass are modeled using a combination of analytical and numerical
results. Although the massive DM halos are identified as galaxy clusters, their masses are not directly observable. It is necessary to use observable quantities, such as the X-ray temperature~\cite{Vikhlinin2009} or SZ flux decrement~\cite{Vanderlinde2010}, to obtain mass-observable scaling relations (for a review, see~\cite{Giodini2013}). The combination of a mass proxy with a theoretical model can be used to predict the number density of galaxy clusters. Therefore, the success of cluster cosmology relies on a good description of cluster counts and clustering, both based on a cosmological model and high-quality observations.

It is known that uncertainties arising from both mass-observable relations and photometric redshifts degrade constraints on cosmological parameters~\cite{Sahlen2009, Shimon2013}. 
These uncertainties comprehend both statistical and systematic errors. 
Other sources of uncertainties are sample-variance and shot-noise (due to the discreteness of the clusters). Optical surveys provide large catalogs (tens of thousand clusters) and, therefore, the shot-noise contribution is negligible with respect to sample-variance, which dominates along with mass uncertainties. On the other hand, X-ray and SZ catalogs contain only tens to hundreds objects. However, they usually have better mass estimates, higher completeness and purity, for example. In this case the cosmological analyses are shot-noise dominated.

Usually the cluster likelihood function is built taking into account the error sources mentioned above. 
Nevertheless, it is assumed that cosmological parameter estimators do not yield extra uncertainties.
The Maximum Likelihood (ML) method is extensively used to estimate the values of parameters given a 
data set. ML estimators are usually consistent, i.e., the expected value of a parameter tends to its 
true value when the size of the data sample ($n$) is sufficiently large, and asymptotically efficient, 
i.e., the variance of the estimator attains the minimum variance bound as $n \rightarrow \infty$. However, ML estimators are not necessarily unbiased and, even if they are consistent, in practical cases the data sample sizes may not be large enough to guarantee that the asymptotic limit is reached~\cite{Lupton1993, Cowan1998, Barlow1999, Bierens2005}.\footnote{A common example is the variance ML estimator  of a Gaussian distribution with mean $\bar{x}$, $\sigma^2_{ML} = \frac{1}{N} \sum_{i = 1}^{N} (x_i - \bar{x})^2$,  which is biased. The unbiased estimator is given by $\sigma^2 = \frac{1}{(N - 1)} \sum_{i = 1}^{N} (x_i - \bar{x})^2$.}

In this work, we study some cosmological parameter estimators in the context of SZ surveys, specifically from the SPT. For the redshift and mass ranges of SPT catalogs, sample variance is negligible~\cite{Hu2003, Vikhlinin2009}. Therefore, we build the likelihood only for the cluster number counts and do not include the spatial clustering. We introduce a formalism to build this likelihood and show that it is similar to the one commonly used in the literature, in the limit where there is only one object in each bin of redshift and mass.  To study the estimators, we use the Monte Carlo approach which consists on randomly sampling artificial data sets from a distribution numerous times. 

To perform this work it was necessary to develop fast and accurate codes. These were implemented in the \textit{Numerical Cosmology} Library (NumCosmo)~\cite{Vitenti2012c} framework, which is open source and adaptable to new models and data, including several cosmological observables.

This paper is organized as follows. In section~\ref{sec:counts} we present the model and assumptions 
used to compute the halo number counts. In section~\ref{sec:likelihood}, we introduce an unbinned likelihood for this observable and compare it with the standard one used in the literature. Then we describe the methodology to compute the bias of an estimator (section~\ref{sec:compute_bias}) and apply it using the likelihood for DM halo abundance (section~\ref{sec:no:uncert}). 
In section~\ref{sec:cl_mass} we study the biases on the cosmological parameter estimators using a SZ mass-observable relation and investigate their dependence on the combination of free parameters fitted and the survey area and depth. We also include the photometric redshift uncertainty. 
In section~\ref{sec:cluster:distpriors} we analyse cosmological parameter estimators obtained from a joint likelihood of cluster number counts and cosmic microwave background data. Finally, in section~\ref{sec:stat_methods}, we compare some estimator error bars computed using three different statistical methods: Monte Carlo, Fisher matrix and profile likelihood. Our concluding remarks are presented in section~\ref{sec:conclusions}.

\section{Cluster number counts}
\label{sec:counts}

The mean number of dark matter halos with mass in 
the range $[M, M + dM]$ and in the redshift interval $[z, z + dz]$ is given by
\begin{equation}\label{dN_dz}
\frac{d^2N}{dz d\ln M} dzd\ln M =  \frac{dV}{dz} \frac{dn(M, z)}{d\ln M} dzd\ln M,
\end{equation}
where $\frac{dn (M, z)}{d\ln M}$ is the comoving number density of halos at 
redshift $z$ with mass $M$ (i.e., the mass function). Assuming a homogeneous and isotropic universe and, as indicated by current observational data, spatially flat, the comoving volume element is
\begin{equation}
\frac{dV}{dz} = \Delta\Omega \left(\frac{\pi}{180}\right)^2 \frac{c}{H(z)} \left(\int_0^z dz^{\prime} \frac{c}{H(z^{\prime})}\right)^2,
\end{equation}
where $\Delta\Omega$ is the survey area in square degrees.
In this paper, we assume a constant DE equation of state parameter $p_{DE}/\rho_{DE} = w_0$, such that the Hubble function is
\begin{equation}\label{eq:hubble}
H(z) = H_0 \big[\Omega_r(1 + z)^4 + \Omega_m(1 + z)^3 + \Omega_{DE} \, (1 + z)^{3(1 + w_0)} \big]^{1/2},
\end{equation}
where $H_0$ is the Hubble constant. The energy density parameters $\Omega_x$ are the ratio 
between the energy density $\rho_x(z = 0)$ and the critical energy density $\rho_{crit} = \frac{3 H_0^2}
{8 \pi G}$. In eq.~\eqref{eq:hubble} $\Omega_r$, $\Omega_m$, and $\Omega_{DE}$ are the radiation, 
matter, and DE density parameters, respectively. As we are assuming a flat universe, 
$\Omega_{DE} = 1 - \Omega_m - \Omega_r$.

The mass function can be written as~\cite{Press1974, Bond1991, Sheth1999} 
\begin{equation}\label{eq:mass_function}
 \frac{dn(M, z)}{d\ln M} = -\frac{\rho_m(z)}{M} f(\sigma_{R}, z) \frac{1}{\sigma_{R}} \frac{d\sigma_{R}}{d\ln M},
\end{equation}
where $\rho_m(z)$ is the mean matter density at redshift $z$, $f(\sigma_{R}, z)$ 
is the so called multiplicity function, which contains information about the nonlinear regime of 
halo formation, and $\sigma^2_{R}$ is the variance of the linear density contrast 
filtered on the length scale $R$ associated to the mass $M$:
\begin{equation}\label{eq:var:R:z}
\sigma^2_R(z) = \int_0^\infty \frac{dk}{2\pi^2} k^2 \, P(k, z) \vert W(k,R) \vert^2,
\end{equation}
where $W(k, R)$ is the window function and $P(k, z)$ is the linear power spectrum. In particular, we use the spherical top-hat function whose Fourier transform is 
\begin{equation}\label{eq:window}
W(k, R) = \frac{3}{(kR)^3} (\sin kR - (kR)\cos kR),
\end{equation}
and encloses a mass $M = 4\pi \rho_m R^3/ 3$.

The linear power spectrum is given by
\begin{equation}\label{eq:powspec}
P(k, z) = A \, k^{n_s} \, T(k)^2 \, D(z)^2, 
\end{equation}
where $n_s$ is the spectral index, $D(z)$ is the linear growth function, and $T(k)$ is the transfer function. 
In this paper, we use the Eisenstein-Hu (EH)~\cite{Eisenstein1998} fitting function for $T(k)$.\footnote{We have checked that using the transfer function from CAMB~\cite{Lewis2000} changes the expected number of DM halos by at most $1\%$ within the redshift and mass ranges considered here. This difference has a negligible impact on the results of this work. The choice for using the EH transfer function is necessary for optimizing the CPU execution time of the Monte Carlo analyses (as described in section~\ref{sec:compute_bias}).}
The power spectrum normalization $A$ may be written in terms of the standard deviation of the density contrast at a scale $R = 8h^{-1}\text{Mpc}$, $\sigma_8$:
\begin{equation}\label{eq:norma_sigma8}
A = \frac{\sigma_8^2}{\int_0^\infty \frac{dk}{2\pi^2} k^{(n_s + 2)} T(k)^2 W^2(k,8)}.
\end{equation}

In this work we compute $D(z)$ by numerically solving the differential equation for the growing mode of linear perturbations~\cite{Uzan2007, Nesseris2008},
\begin{equation}\label{eq:growth}
\frac{d^2D}{dz^2} - \left( \frac{1}{1+z} - \frac{1}{E(z)} \frac{dE(z)}{dz} \right) \frac{dD}{dz} = \frac{3\Omega_m}{2E^2(z)} (1+z) D,
\end{equation}
where we defined the normalized Hubble function as $E(z) = H(z)/H_0$ and $D(z)$ is normalized to unity at $z = 0$.

The first functional form for $f(\sigma)$ was obtained analytically by Press \& Schechter assuming the spherical collapse of initial density peaks~\cite{Press1974}. The multiplicity function was shown to be {\it universal}, in the sense that it does not depend on the cosmological parameters nor the initial conditions. A semi-analytic expression for $f(\sigma)$ was obtained by Sheth \& Tormen~\cite{Sheth1999} considering an ellipsoidal collapse approximation and a statistical treatment using a moving barrier. Their results were shown to be in broad agreement with cosmological N-body simulations. This motivated multiplicity functions to be fitted directly from the simulations~\cite{Jenkins2001, Warren2006, Tinker2008}, providing more accurate fitting functions. In addition, the universality property was verified to a good approximation in the simulations. In this work, we use the Tinker et al.~\cite{Tinker2008} multiplicity function given by
\begin{equation}\label{eq:mult}
f(\sigma_R(z), z) = A \left[ \left( \frac{\sigma_R(z)}{b} \right) ^{-a} + 1 \right] \exp(-c\sigma_R^{-2}(z)),
\end{equation}
where the values of the parameters $A$, $a$, $b$, and $c$ depend on the redshift and halo mass definition.
To allow for a more direct comparison with SZ data, we consider the mass contained in a spherical region 500 times the critical density at $z$, i.e., $R_{\Delta} = \left(3 M_{\Delta} / 4\pi \Delta \rho_{crit}\right)^{1/3}$, with $\Delta = 500$. Reference~\cite{Tinker2008} provides values for the mass function 
parameters for various values of $\Delta$. Following their recommendation, we use spline interpolation to obtain these parameters for $\Delta = 500$ to a good accuracy.
It is worth noting that, in our analyses, we neglect uncertainties related to the determination of the multiplicity function. In particular, we ignore the errors on the fitted parameters $A$, $a$, $b$ and $c$. 
In reference~\cite{Cunha2010} the authors show that while errors of the order of $10\%$ in the mass function degrade the constraints on cosmological parameters, this effect is small when compared with that of the mass-observable relation. 

\section{Likelihood Model}
\label{sec:likelihood}

In this section, we review the standard likelihood employed in the literature for halo abundance 
and introduce one based on the Extended Maximum Likelihood method.\footnote{In this section, we refer to DM halos instead of galaxy clusters because later we will introduce a mass proxy to properly build the likelihood for cluster abundance in terms of a cluster observable.}
We show that these two likelihoods are equivalent under some conditions. The later will be employed to
check if the cosmological parameter estimators are biased (for the considered sample sizes).

\subsection{Poisson Distribution}
\label{sec:poisson}

The standard likelihood of the mean
halo number counts is built assuming a Poisson distribution in bins of redshift and mass~\cite{Cash1979, Borgani2001, 
Holder2001, Vikhlinin2009, Vanderlinde2010, Rozo2010, Sehgal2011, Benson2013}, namely, 
\begin{equation}\label{eq:poisson}
\mathcal{L} (\{\lambda_k(\{\theta_j\})\},\{n_k\}) = \prod_{k = 1}^{\mathcal{N}} \frac{\lambda_k^{n_ 
k}e^{-\lambda_k}}{n_k!},
\end{equation}
where $\mathcal{N}$ is the number of bins in the $\ln M-z$ space (or in the \textit{mass-observable} - $z$ 
space), $n_k$ is the observed number of halos (clusters) in the $k$-th bin and the random variable 
$\lambda_k(\{\theta_j\})$ is the expected number of halos in the $k$-th bin for a given model.

In the regime where sample variance is negligible, a common practice is to
take the limit where there is at most one halo per bin. In this case eq.~\eqref{eq:poisson} takes the form 
\begin{align}
\mathcal{L} (\{\lambda_k(\{\theta_j\})\},\{n_k\}) &= \prod_{k = 1}^{n} \lambda_k e^{-\lambda_k} 
\prod_{l \neq k} e^{-\lambda_k} \nonumber \\
&= \prod_{k = 1}^{n} \lambda_k \prod_{all \, bins} e^{-\lambda_k}, 
\end{align}
where now the first product is over the halos in the sample. The log likelihood is thus
\begin{equation}\label{eq:ln_poisson}
\ln \mathcal{L} (\{\theta_j\}, \{M_i, z_i\}) = \sum_{k = 1}^{n} \left[ \ln (\lambda_k(\{\theta_j\})) 
\right] - N(\{\theta_j\}), 
\end{equation}
where $\lambda_k(\{\theta_j\}) = \frac{d^2N(M_i, z_i, \{\theta_j\})}{dz d\ln M}$ (e.g.~\cite{Borgani2001, Sehgal2011}),  $N (\{\theta_j\})$ is the predicted number of halos with redshift and mass values within, respectively, $[z_{min}, z_{max}]$ and $[M_{min}, M_{max}]$, and $\{\theta_j\}$ represents the set of cosmological parameters (it may also include other parameters such as those from a 
mass-observable relation).

\subsection{Extended Maximum Likelihood}\label{sec:halo_likelihood}

In this work, we introduce an unbinned likelihood \cite{Lima2010} which we build following the extended maximum 
likelihood (EML) method~\cite{Cowan1998, Barlow1999}. This approach does not require a normalized 
probability distribution, i.e., the number of events is not fixed. Particularly, the likelihood 
is the product of two terms. The first is the Poisson probability to find $n$ halos in the whole mass and redshift ranges of the sample. The second 
term is the product of $n$ terms, each one given by  the probability distribution of a 
halo  to have mass in the range $[M, M + dM]$ and to be found in the redshift interval $[z, z + dz]$, 
namely,
\begin{equation}\label{eq:pmz_dis}
P(M, z, \{\theta_j\}) dz dM = \frac{d^2N(M, z, \{\theta_j\})}{dzd\ln M}\frac{ dz d\ln M}{N (\{\theta_j\})},
\end{equation}
where $N (\{\theta_j\})$ is the normalization factor.

Thus the likelihood function is
\begin{equation}
\mathcal{L} (\{\theta_j\}, \{M_i, z_i\}) = \frac{N^{n}e^{- N}}{n!} 
\prod_{i=1}^{n} \frac{1}{N} \frac{d^2N(M_i, z_i, \{\theta_j\})}{dz d\ln M},
\end{equation}
where $z_i$ is the redshift and $M_i$ is the the mass of each halo. Therefore the log likelihood is  
\begin{equation}\label{eq:ln_eml}
\ln \mathcal{L} (\{\theta_j\}, \{M_i, z_i\}) = \sum_{i=1}^{n} \ln \left( \frac{d^2N(M_i, z_i, \{\theta_j\})}{dz d\ln M} \right) - N (\{\theta_j\}) - \ln n!.
\end{equation}
Note that dropping the last term of eq.~\eqref{eq:ln_eml}, since it does not depend on $\{\theta_j\}$, 
lead us to eq.~\eqref{eq:ln_poisson}. In this case the EML and Poisson methods are identical. 

One advantage of the EML is the possibility to consider different informations for each cluster (such as mass-observable relations from different observables), since the likelihood is naturally a product of individual probabilities for each cluster. Also, the likelihood is naturally unbinned, such that it is free from possible bias that could be associated to the choice of the bin size. For this reason we will use the EML throughout this work. It is worth noting that, in ref.~\cite{Mantz2010}, the authors also build an unbinned likelihood using the truncated regression approach. Their likelihood is equivalent to eq.~\eqref{eq:ln_eml} except for a Poisson term of the number of missing objects $N_{mis}$, which is equal to unity when the likelihood is marginalized over $N_{mis}$.

\section{Obtaining the bias of cosmological parameter estimators with the Monte Carlo method}
\label{sec:compute_bias}

The estimators for a given set of parameters ($\{\hat{\theta}_j\}$) are obtained by maximizing the likelihood with respect to these parameters.
In order to analyse whether the estimators obtained from a likelihood function 
are biased, we need to calculate their expected values given a fiducial model 
$\{\theta^ 0_j\}$. The bias of $\hat{\theta}_j$ is defined as
\begin{equation}
\label{eq:bias}
b_{\hat{\theta}_j} = \langle \hat{\theta}_j \rangle - \theta^0_j.
\end{equation} 

The expected value of the cosmological parameter estimators cannot be computed analytically from the cluster number counts likelihood. To obtain $\langle \hat{\theta}_j \rangle$ we employ the Monte Carlo (MC) method, i.e., we generate a set of realizations of the cluster distribution in mass and redshift from a given input fiducial model and compute the best-fitting value of $\hat{\theta}_j$ for each realization by minimizing the likelihood function ($-2\ln \mathcal{L}$) [eq.~\eqref{eq:ln_eml}] with respect to $\theta_j$. Assuming the probability distribution given by eq.~\eqref{eq:pmz_dis}, each realization is a catalog of DM halos containing their redshifts $z_i$ and masses $M_i$ (see appendix~\ref{ap:sampling} for a detailed description). Then we compute the expected value using as an estimate the arithmetic mean $\overline{\hat{\theta}}_j$, i.e.,
\begin{equation}
\label{eq:expec_val}
\langle \hat{\theta}_j \rangle = \overline{\hat{\theta}}_j = \sum_{l = 1}^{m} \frac{\hat{\theta}_{jl}}{m},
\end{equation}
where $m$ is the number of realizations and $\hat{\theta}_{jl}$ is the best-fitting value for the l-th 
realization.  

The variance of $\overline{\hat{\theta}}_j$ is given by
\begin{equation}\label{eq:sd_bias2}
\sigma^2 \left(\overline{\hat{\theta}}_j \right) = \frac{\sigma^2\left(\hat{\theta}_j\right)}{m},
\end{equation}
where $\sigma^2 \left(\hat{\theta}_j \right)$ is the variance of the estimator $\hat{\theta}_j$. In cases where $b_{\hat{\theta}_j} \neq 0$, the significance of this bias is ensured by computing eq.~\eqref{eq:sd_bias2}. For example, if the estimator has a standard deviation of $1\%$ in $\hat{\theta}^0_j$ and $b_{\hat{\theta}_j}$ is $5\%$ of $\hat{\theta}^0_j$, then we must generate 100 realizations such that $ \sigma \left(\overline{\hat{\theta}}_j\right) = 0.1\%$ is one order smaller than $b_{\hat{\theta}_j}$. Note that this procedure usually requires a large number of realizations. In particular for our purposes, it is necessary to generate 10,000 realizations for each case, i.e., each choice of the fiducial model.

Therefore, to make this work feasible, it was necessary to develop numerical codes not only fast but whose framework could concatenate the different steps, i.e., generate a realization, build the likelihood, and perform the statistical analysis. Those algorithms are implemented in the NumCosmo library~\cite{Vitenti2012c}.

\section{Bias from halo abundance likelihood}
\label{sec:no:uncert}

In this section, we apply the methodology described above to compute the bias of some cosmological parameter estimators in an idealized scenario where both the redshift and mass are perfectly known. For this, we have to adopt a fiducial model, which comprises both cosmological and survey parameters. Our choice is based on SZ surveys, since sample variance is usually negligible for their configurations. 
First, we assume a constant minimum mass threshold for the entire redshift interval. This is a good approximation for SZ surveys for which the mass limit for detection $M_{min}$ is nearly redshift  independent and is weakly dependent on cosmology~\cite{Carlstrom2000, Holder2000, Haiman2001}. Specifically, we set $M_{min} = 2 \times 10^{14} h^{-1} M_\odot$, which corresponds to the minimum value of the SZ observable that we will describe in section~\ref{sec:cl_mass}.

Besides the minimum mass threshold, we define a fiducial model by the redshift interval, the solid angle, and the set of cosmological parameters $\{\theta_j\}$. They are, respectively, $[0.3, 1.1]$, $\Delta\Omega = 720 \text{deg}^2$ and
\begin{equation}
\label{eq:set_cosmo_params}
\begin{split}
\{\theta_j\} &= \{\Omega_c, \Omega_b, H_0, n_s, \sigma_8, w_0\}\\
= &\{0.244, 0.0405, 73.9, 0.966, 0.766, -1\},\\
\end{split}
\end{equation}
where $\Omega_c$ is the cold dark matter density, $\Omega_b$ is the baryonic matter density 
($\Omega_m = \Omega_c + \Omega_b$). 
The reason for choosing these specific values will become clear later in section~\ref{sec:results}. 

Having defined the methodology to compute the bias of an estimator and the fiducial model, we now use them to determine whether the estimators are biased or not and if the biases obtained are relevant for cosmological studies. Throughout this paper, we focus our analyses on the estimators for $\Omega_c$, $\sigma_8$, and $w_0$,  since the first two are the cosmological parameters better constrained by cluster abundance (see, e.g.,~\cite{Holder2001, Haiman2001}) and the later is the parameter used to probe the nature of dark energy, whose understanding is a major goal of ongoing and future surveys.

In this section, we study how the estimators' bias size vary with respect to the number of free parameters fitted. For this sake, we compute their biases fitting the respective estimators in one, two, and three-dimensional parametric spaces, i.e., $(\Omega_c)$, $(\sigma_8)$, $(w_0)$, $(\Omega_c, \sigma_8)$, $(\Omega_c, w_0)$, $(\sigma_8, w_0)$, and $(\Omega_c, \sigma_8, w_0)$. The remaining parameters (cosmological and from the survey configurations) 
are always kept fixed.

Throughout this paper we report the bias on a parameter as a percentage of the $1\sigma$ error bar on this parameter. This helps to assess the significance of the bias in each case, i.e., even if the bias is large when compared to the estimated parameter values, it can be irrelevant if the error on this estimate is much larger. We define the following notation for the relative biases $B_{\hat{x}} \equiv b_{\hat{x}}/\sigma(\hat{x})$, i.e., $$B_{\hat{\Omega}_c} \equiv \frac{b_{\hat{\Omega}_c}}{\sigma({\hat{\Omega}_c})},\quad B_{\hat{\sigma}_8} \equiv \frac{b_{\hat{\sigma}_8}}{\sigma({\hat{\sigma}_8})},\quad B_{\hat{w}_0} \equiv \frac{b_{\hat{w}_0}}{\sigma({\hat{w}_0})},$$
where $b_{\hat{x}}$ is the bias of the estimator $\hat{x}$ computed using eqs. \eqref{eq:bias} and \eqref{eq:expec_val}, and $\sigma(\hat{x})$ is the standard deviation of $\hat{x}$. 

For each of the 7 parametric spaces defined above, we perform a MC with 10,000 realizations for the chosen fiducial model. This configuration provides an expected number about 80 clusters. The results 
for the mean values of the parameters, their variance, and their bias are shown in table~\ref{tab:720deg2:nou}. 
For a single free parameter the three estimators have small relative biases (about $2\%$). On the other hand, estimators in the bi-dimensional parametric spaces may have larger relative biases.
For example, in the plane $(\Omega_c, w_0)$,  $\hat{\Omega}_c$ has a relative bias of $23\%$. We also obtain a significant relative bias of $20\%$ for $\hat{\sigma}_8$ when we fit this parameter together with $w_0$. For $(\Omega_c, \sigma_8)$, both estimators have small relative biases.
In the three-dimensional parametric space, $\hat{\Omega}_c$ and $\hat{w}_0$ have small biases while $\hat{\sigma}_8$ has a relative bias of $37\%$. This case evinces the importance of knowing the properties of the cosmological parameter estimators. 

\begin{table*}
\begin{center}
\begin{tabular}{|c|c|c|c|c|c|c|}
\multicolumn{7}{c}{DM halo abundance likelihood, $\Delta \Omega = 720\text{deg}^2$} \\
\hline
Free & $\bar{\hat{\Omega}}_c \pm \sigma(\hat{\Omega}_c)$ & $B_{\hat{\Omega}_c}$ & $\bar{\hat{\sigma}}_8 \pm \sigma(\hat{\sigma}_8)$ & $B_{\hat{\sigma}_8}$ & $\bar{\hat{w}}_0 \pm \sigma(\hat{w}_0)$ & $B_{\hat{w}_0}$ \\ \hline
1 & $0.244 \pm 0.011$ & $3\%$ & $0.766 \pm 0.006$ & $2\%$ & $-1.00 \pm 0.13$ & $2\%$ \\
\hline
2 & $0.246 \pm 0.033$ & $ 6\%$ & $0.766 \pm 0.020$ & $2\%$ & $--$ & $--$ \\
2 & $0.253 \pm 0.043$ & $ 23\%$ & $--$ & $--$ & $-0.94 \pm 0.46$ & $ 12\%$ \\
2 & $--$ & $--$ & $0.770 \pm 0.022$ & $ 20\%$ & $-0.98 \pm 0.28$ & $ 8\%$ \\
\hline
3 & $0.239 \pm 0.077$ & $ 6\%$ & $0.777 \pm 0.029$ & $37 \%$ & $-1.00 \pm 0.63$ & $ < 1\%$\\
\hline
\end{tabular}
\caption{\label{tab:720deg2:nou} First Column: Number of fitted parameters. Even Columns: Estimated expected value of $\hat{x}$ and its standard deviation $\sigma(\hat{x})$ obtained fitting simultaneously 1 (first row), 2 (second, third, and forth rows), and 3 (last row) cosmological parameters and keeping the others fixed. Odd Columns: Bias size $b_{\hat{x}}$ relative to $\sigma(\hat{x})$.}
\end{center}
\end{table*}

\section{Biases from cluster counts including a mass-observable relation}
\label{sec:cl_mass}

So far we presented a likelihood, a methodology, and a first analysis considering only DM halos.
The relatively large biases found in the example above, in particular the significant bias obtained for  ${\hat{\sigma}}_8$ in the case of 3 free parameters, pushes us to consider more realistic situations.
A fundamental ingredient for modelling cluster abundance is to consider 
a mass proxy and associated mass-observable relation.
Here we extend the formalism of the preceding sections to consider specifically the mass-observable relation for SZ surveys and take the fiducial values from SZ catalogs. We study how the relative biases change as a function of the number of parameters fitted and the size of the sample (changing both the area of the survey and the redshift depth). 

\subsection{Sunyaev--Zel'dovich mass-observable relation}

One way to detect galaxy clusters and estimate their masses is through the SZ effect, which consists of a distortion in the frequency distribution of the cosmic microwave background photons due to inverse Compton scattering by the hot intracluster medium. The magnitude decrement due to the SZ effect (or increment for photons with frequencies $\gtrsim 217$ GHz) is a proxy for the mass of the cluster (see~\cite{Giodini2013} and references therein).  

In this work, we use the mass-observable relation developed by the SPT team in refs.~\cite{Vanderlinde2010, Benson2013, Reichardt2013}, where the detection significance $\xi$ is used as a SZ mass proxy~\cite{Staniszewski2009}. In ref.~\cite{Vanderlinde2010} the authors obtained, using simulations, an unbiased significance $\zeta$ such that 
\begin{equation}\label{eq:xi_to_zeta}
\zeta = \sqrt{\langle\xi\rangle^2 - 3},
\end{equation}
and $\xi$ is related to $\langle\xi\rangle$ by a Gaussian scatter of unit width, to which we will refer as $P(\xi | \zeta)$.

The scaling relation between $\zeta$ and $M_{500}$ is given by~\cite{Benson2013, Reichardt2013}
\begin{equation}\label{eq:zeta_to_mass}
\zeta = A_{SZ} \left( \frac{M_{500}}{3 \times 10^{14} M_\odot h^{-1}} \right)^{B_{SZ}} \left( \frac{E(z)}{E(0.6)} \right)^{C_{SZ}},
\end{equation}
where $A_{SZ}$ is the normalization, $B_{SZ}$ is the slope, and $C_{SZ}$ is the redshift evolution parameter. Following refs.~\cite{Benson2013, Reichardt2013}, we assume that $P(\ln\zeta | \ln M)$ is a Gaussian distribution with scatter given by $D_{SZ}$.

In section~\ref{sec:halo_likelihood} we built the likelihood for halo abundance. Analogously, given the $\xi - M_{500}$ relation, we have that the likelihood for cluster number counts is
\begin{equation}\label{eq:ln_eml:unbinned}
\ln \mathcal{L} (\{\theta_j\}, \{\xi_i, z_i\}) = \sum_{i=1}^{n} \ln \left( \frac{d^2N(\xi_i, z_i, \{\theta_j\})}{dz d\xi} \right) - N (\{\theta_j\}) - \ln n!,
\end{equation}
where 
\begin{equation}\label{eq:d2N_z:xi}
\frac{d^2N(\xi_i, z_i, \{\theta_j\})}{dz d\xi} = \int d\ln M \int d\zeta \frac{d^2N(M, z_i, \{\theta_j\})}{dz d\ln M} \, P(\xi_i | \zeta) \, P(\ln\zeta | \ln M).
\end{equation}
Note that the normalization factor $N (\{\theta_j\})$ is now the expected number of clusters with $z_{min} \leq z \leq z_{max}$ and $\xi \geq \xi_{min}$. We assume that all objects with $\xi \geq \xi_{min}$ are detected in the entire redshift interval, i.e., the catalog is complete above this threshold.

\subsection{Fiducial values from SPT}
\label{sec:results}

Now we shall apply the methodology of section~\ref{sec:compute_bias} extending our analyses to include the SZ mass-observable relation, considering realistic parameters values taken from SZ surveys. The MC approach is performed by minimizing eq.~\eqref{eq:ln_eml:unbinned} with respect to $\theta_j$, assuming the probability distribution given by eq.~\eqref{eq:pmz_dis} now in terms of $\xi$ and the mass-observable relations $P(\xi| \zeta)$ and $P(\ln\zeta| \ln M)$. Each realization is now a catalog of clusters containing their redshifts $z_i$ and the detection significance $\xi_i$ (see appendix~\ref{ap:sampling} for a discussion on the sampling).

In this case, the fiducial model used to generate the realizations also includes the parameters of the $\zeta - M_{500}$ relation and the minimum value of the  detection significance $\xi_{min}$. Thus, along with  the cosmological parameters, we have
\begin{equation}
\begin{split}
\{\theta_j\} = \{&\Omega_c, \Omega_b, H_0, n_s, \sigma_8, w_0, A_{SZ}, B_{SZ}, C_{SZ}, D_{SZ}\} \\
= &\{0.244, 0.0405, 73.9, 0.966, 0.766, -1, 5.31, 1.39, 0.9, 0.21\}, \\
\end{split}
\end{equation}
and $\xi_{min} = 5.0$. The redshift interval, $[0.3, 1.1]$, and the threshold significance $\xi_{min}$ are taken from the SPT catalog given in ref.~\cite{Benson2013}. The parameters $\{\theta_j\}$ correspond to the best-fit values obtained in this same reference by combining cluster abundance, Hubble parameter, and Big-Bang nucleosynthesis measurements, assuming a $\Lambda$CDM model. 

Similarly to the analyses carried out in section~\ref{sec:no:uncert}, we compute the relative biases of $\hat{\Omega}_c$, $\hat{\sigma}_8$, and $\hat{w}_0$ in one, two, and three-dimensional parametric spaces. In section~\ref{sec:no:uncert} we fixed  
$\Delta \Omega = 720 \, \text{deg}^2$, which is the area covered by the SPT catalog from ref.~\cite{Reichardt2013}. In this section we also consider 
$\Delta \Omega = 178$~\cite{Benson2013} and $2500 \, \text{deg}^2$. The 
later being the total area covered by the SPT Survey~\cite{Story2013}. Therefore, we are able to check how the estimator biases vary with the number of cosmological parameters being fitted simultaneously as well as with the size of the catalogs. As in section~\ref{sec:no:uncert}, all other parameters are kept fixed in the following analyses (including those from the mass-observable relation) and in each case we perform a MC with 10,000 realizations. 

\subsection{Dependence on the number of parameters fitted}
\label{ParamSpace}

\begin{figure*}
\includegraphics[scale=0.4]{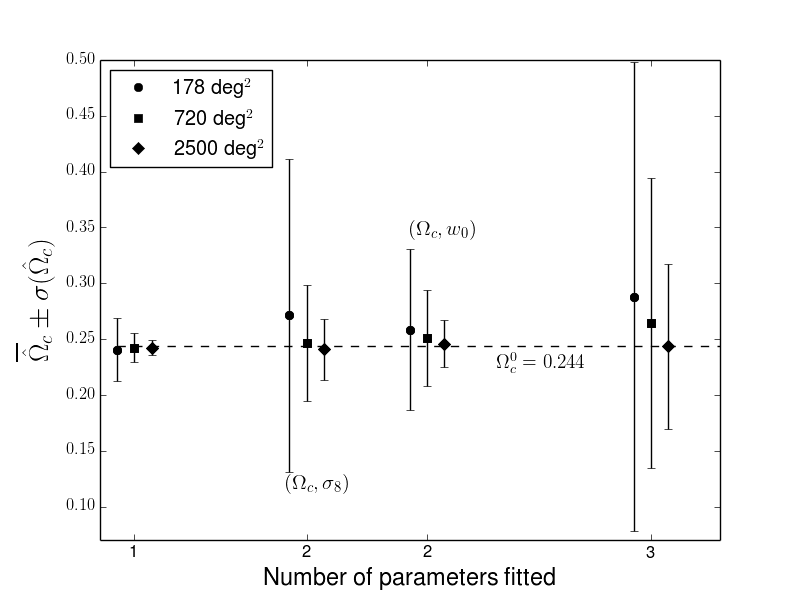}
\includegraphics[scale=0.4]{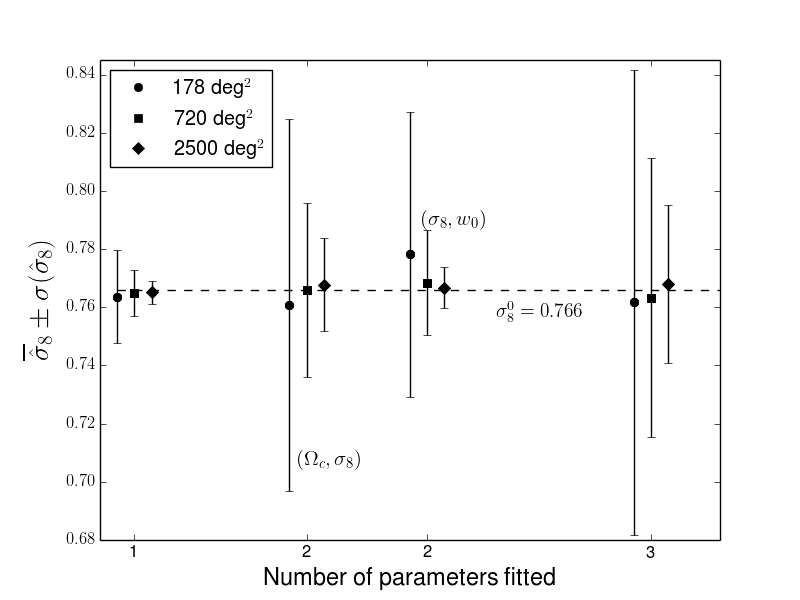}
\begin{center}
\includegraphics[scale=0.4]{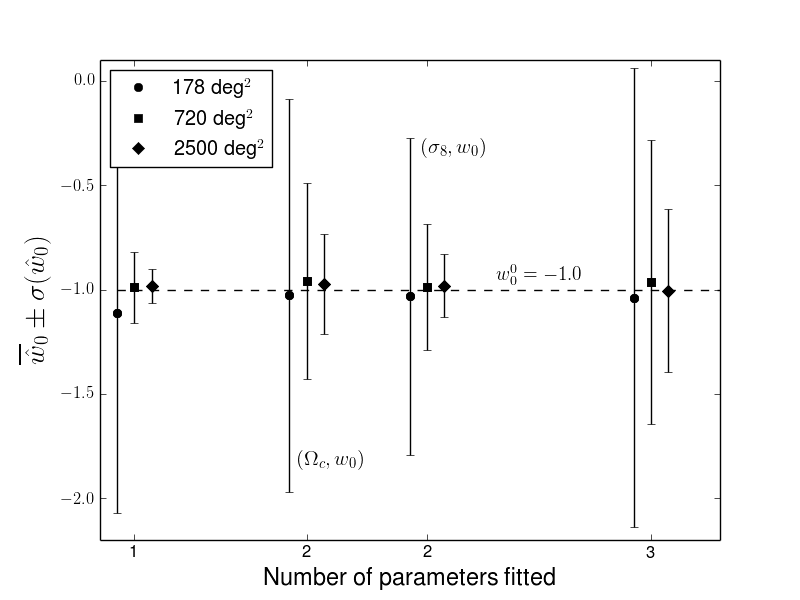}
\end{center}
\caption{Expected values and error bars of the estimators $\hat{\Omega}_c$ (upper left panel), $\hat{\sigma}_8$ (upper right panel) and $\hat{w}_0$ (lower panel) computed for three different survey areas and the four parametric spaces probed. The dashed lines show the fiducial values.}
\label{fig:nparams}
\end{figure*}

The results for the several combinations of the parameters being fitted and the 3 values of the area discussed in the previous section are displayed in figure~\ref{fig:nparams}, which shows that the bias of the estimators are only a relatively small fraction of the size of the error bars. 

In the one-dimensional parametric spaces 
the relative biases are generally greater than $10\%$ but smaller than $25\%$. 
For both $\hat{\Omega}_c$ and $\hat{\sigma}_8$ the biases, $b_{\hat{\Omega}_c}$ and $b_{\hat{\sigma}_8}$, decrease with the increase of the survey area, as expected.  
Nevertheless, the errors bars $\sigma(\hat{\Omega}_c)$ and $\sigma(\hat{\sigma}_8)$ have a steeper decrease, such that relative biases $B_{\hat{\Omega}_c}$ and $B_{\hat{\sigma}_8}$ end up increasing with the area. This shows that even for large samples, where one would expect the biases related to the cluster abundance likelihood to be smaller, this bias could still be a significant fraction of the error bars. 

The freedom introduced by the increased number of fitted parameters results in larger error bars in almost all estimators, as expected. 
In general, the absolute biases $\vert b_{\hat{x}} \vert$ also increase with the number of fitted parameters. 
For the two and three-dimensional parametric spaces the relative biases, $B_{\hat{\Omega}_c}$ and $B_{\hat{w}_0}$ are roughly constant for both $178$ and $720 \, \text{deg}^2$. Namely, $B_{\hat{w}_0} \lesssim 10\%$ and $B_{\hat{\Omega}_c} \simeq 20\%$. On the other hand, $B_{\hat{\sigma}_8} \lesssim 10\%$ for the $(\Omega_c, \sigma_8)$ and $(\Omega_c, \sigma_8, w_0)$ spaces (being $< 1\%$ for the former space and $720 \, \text{deg}^2$), while $B_{\hat{\sigma}_8} \simeq 25\%$ and $14\%$ in the plane $(\sigma_8, w_0)$ for 178 and $720 \, \text{deg}^2$, respectively.

Only for $\Delta\Omega = 2500 \, \text{deg}^2$ we note a consistent decrease of the relative biases with the number of parameters fitted. In this case, all estimators present values for $B_{\hat{x}}$ of the order of $20\%$, $10\%$, and $< 10\%$ for one, two and three-dimensional parametric spaces, respectively.

\subsection{Dependence on the survey area and depth}
\label{AreaDepth}

For most cases in the upper panels of figure~\ref{fig:nparams} we see that the absolute values of the bias decrease with the area, as expected from the increase of the sample size. However, one may wonder if this trend continues for higher areas and whether the expected values converge to the fiducial values 
(it is worth mentioning that the sample size corresponding to $2500 \, \text{deg}^2$ is of order of 250 objects). Therefore, we extend the analysis to larger areas and include also some intermediate values to see the general evolution scaling with area. In particular, in addition to the three areas of the previous section we run the MC simulations for $\Delta\Omega = 450, 1600, 3250$, and $4000 \, \text{deg}^2$. The later corresponds to the original planned SPT footprint~\cite{Ruhl2004}. This study is carried out only in the three-dimensional parametric space. 

\begin{figure*}
\includegraphics[scale=0.4]{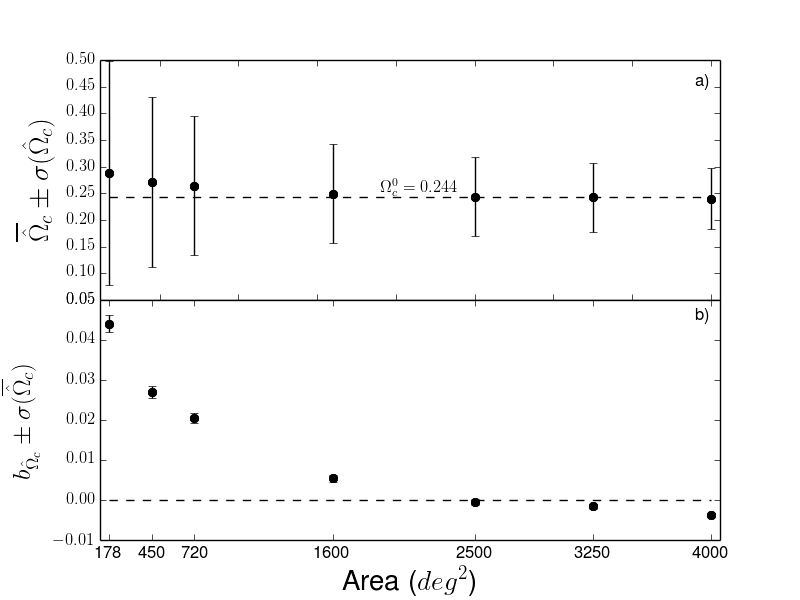}
\includegraphics[scale=0.4]{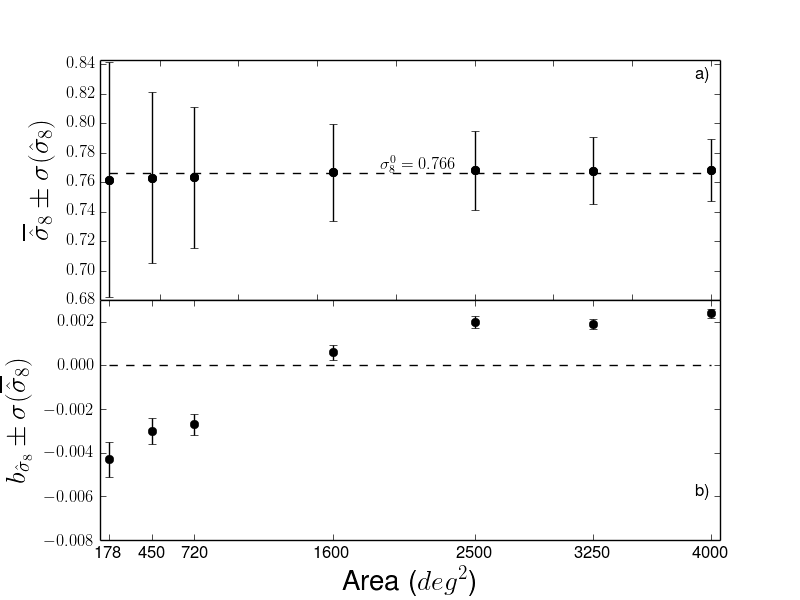}
\begin{center}
\includegraphics[scale=0.4]{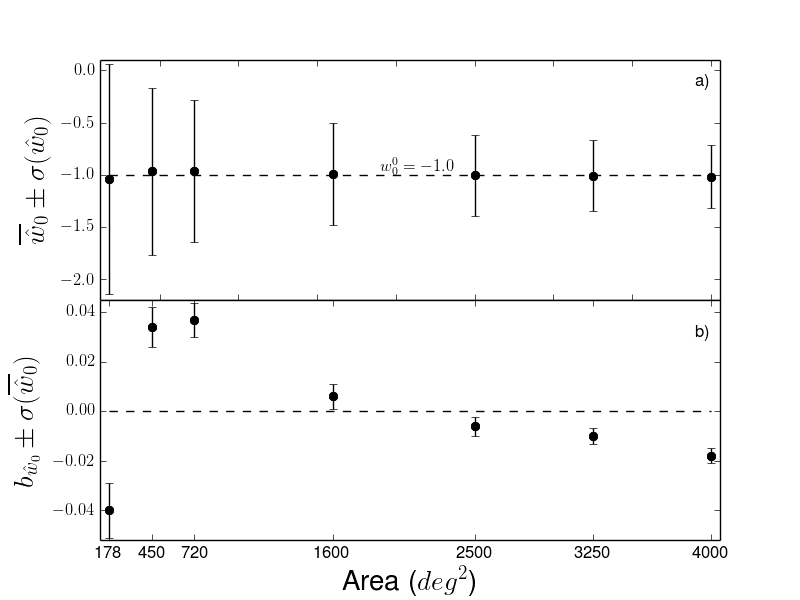}
\end{center}
\caption{The expected values of the estimators (dots) are displayed with the error bars of the estimators (a) and of their means (b) for different survey areas.}
\label{fig:area}
\end{figure*}

Figure~\ref{fig:area} summarizes the results for the seven areas considered. We note that the three parameters present the larger absolute biases for the 3 smaller areas. The estimators are closer to their true values for $1600$ ($\hat{\sigma}_8$) or $2500 \, \text{deg}^2$ ($\hat{\Omega}_c$ and $\hat{w}_0$). Nevertheless, all estimators have their biases increased for the larger areas. 
In addition to displaying the values of the biases and the uncertainties in the parameters being fitted, in this figure we also show the uncertainties associated to the values of $b_{\hat {x}}$, i.e., $\sigma \left(\overline{\hat x} \right)$. We see that the biases are being precisely estimated. 

To better understand the increase of the biases for larger areas, we also compute the estimators of the one- and three-dimensional parametric spaces for 10,000 and $40,000 \text{deg}^2$, being the last close to the full-sky limit. The results are displayed in table~\ref{tab:large_areas}. The biases continue to grow with the increase of the area and, therefore, we have that $\hat{\Omega}_c$, $\hat{\sigma}_8$, and $\hat{w}_0$ are not consistent (i.e. $\Lim{n\rightarrow \infty}b_{\hat{x}} \neq 0$), in the sense that we are treating the full-sky area as the the asymptotic limit. 
We also note that the estimate of the expected value of all estimators computed in the one-dimensional parametric space have already converged for $10,000 \, \text{deg}^2$ (since $\overline{\hat{x}}$'s are equal for both areas), but not to their fiducial values, i.e., $b_{\hat{x}} \neq 0$. It is worth emphasizing that in the full-sky limit the bias is a dominant error source, being the relative biases about $30\%$ to $50\%$ for $(\Omega_c, \sigma_8, w_0)$ and about $80\%$ for the one-dimensional parametric spaces.

\begin{table}
\begin{tabular}{|c|c|c|c|c|c|c|c|}
\multicolumn{7}{c}{$\xi - M_{500}$ cluster abundance likelihood, $z_{max} = 1.1$} \\
\hline
$\Delta \Omega [\text{deg}^2]$ & Free & $\bar{\hat{\Omega}}_c \pm \sigma(\hat{\Omega}_c)$ & $B_{\hat{\Omega}_c}$ & $\bar{\hat{\sigma}}_8 \pm \sigma(\hat{\sigma}_8)$ & $B_{\hat{\sigma}_8}$ & $\bar{\hat{w}}_0 \pm \sigma(\hat{w}_0)$ & $B_{\hat{w}_0}$ \\ \hline
\multirow{2}{1cm}{$10,000$}& 1 & $0.2425 \pm 0.0036$ & $42\%$ & $0.7653 \pm 0.0021$ & $33\%$ & $-0.983 \pm 0.041$ & $42\%$\\
& 3 & $0.2367 \pm 0.0351$ & $ 21\%$ & $0.7690 \pm 0.0133$ & $23 \%$ & $-1.026 \pm 0.186$ & $ 14\%$\\
\hline
\multirow{2}{1cm}{$40,000$} & 1 & $0.2425 \pm 0.0018$ & $83\%$ & $0.7653 \pm 0.0010$ & $70\%$ & $-0.983 \pm 0.021$ & $81\%$ \\
& 3 & $0.2349 \pm 0.0174$ & $ 52\%$ & $0.7694 \pm 0.0067$ & $51 \%$ & $-1.03 \pm 0.09$ & $ 33\%$\\
\hline
\end{tabular}
\caption{\label{tab:large_areas} First Column: Survey area. Second Column: Number of fitted parameters. Odd Columns: Estimated expected value of $\hat{x}$ and its standard deviation $\sigma(\hat{x})$ obtained fitting simultaneously 1 (first and third rows) and 3 (second and forth rows) cosmological parameters and keeping the others fixed. Even Columns: Bias size $b_{\hat{x}}$ relative to $\sigma(\hat{x})$.}
\end{table}

In order to identify the cause of the \emph{not consistent} feature of the cosmological parameter estimators, $\hat{\Omega}_c$, $\hat{\sigma}_8$, and $\hat{w}_0$, we carry out the MC analyses considering $40,000 \, \text{deg}^2$ and two other likelihoods: the halo abundance [given by eq.~\eqref{eq:ln_eml}] and a cluster abundance likelihood given by
\begin{equation}
\ln \mathcal{L} (\{\theta_j\}, \{M^{obs}_i, z_i\}) = \sum_{i=1}^{n} \ln \left( \frac{d^2N(M^{obs}_i, z_i, \{\theta_j\})}{dz d\ln M^{obs}} \right) - N (\{\theta_j\}) - \ln n!,
\end{equation}
where 
\begin{equation}
\frac{d^2N(M^{obs}_i, z_i, \{\theta_j\})}{dz d\ln M^{obs}} = \int d\ln M \frac{d^2N(M, z_i, \{\theta_j\})}{dz d\ln M} \, P(\ln M^{obs} | \ln M),
\end{equation}
and $P(\ln M^{obs} | \ln M)$ is a log-normal probability distribution for obtaining an observable mass $M^{obs}$ given the true mass $M$, 
\begin{equation}
P(\ln M^{obs} | \ln M) \, d\ln M^{obs} = \frac{1}{\sqrt{2\pi}\sigma_{\ln M}} 
e^{- \frac{\left(\ln M^{obs} - \ln M\right)^2}{2\sigma^2_{\ln M}}} \, d\ln M^{obs},
\end{equation}
with $\sigma_{\ln M} = D_{SZ}$.
For both likelihoods, we obtain that all estimators are unbiased, i.e., the fiducial values are recovered within $\sigma\left(\overline{\hat{x}}\right)$, such that the relative biases are less than $1\%$. Therefore, while the halo abundance likelihood provides biased estimators, when the sample size is small (see section~\ref{sec:no:uncert}), but is consistent, the $\xi - M_{500}$ cluster abundance likelihood yields biased estimators whose importance grows with the increase of the sample size. It is worth emphasizing that we cross-checked our code with the SPT team's one and, therefore, we discard the hypothesis that the results obtained with $\xi - M_{500}$ cluster abundance likelihood are due to bugs in the code.\footnote{The likelihood code of the SPT team is available at http://pole.uchicago.edu/public/data/reichardt12/index.html.}

Besides the area, another factor that determines the sample size is the redshift depth. We therefore study the behavior of the bias and the statistical uncertainties as a function of the maximum redshift of the survey. In particular, we perform the MC analyses (using the $\xi - M_{500}$ cluster abundance likelihood) for seven values of the maximum redshift equally spaced in the interval $[0.9, 1.5]$. The later, $z_{max} = 1.5$, corresponds to the highest redshift of the SPT catalog in ref.~\cite{Reichardt2013}. 

\begin{figure*}
\includegraphics[scale=0.4]{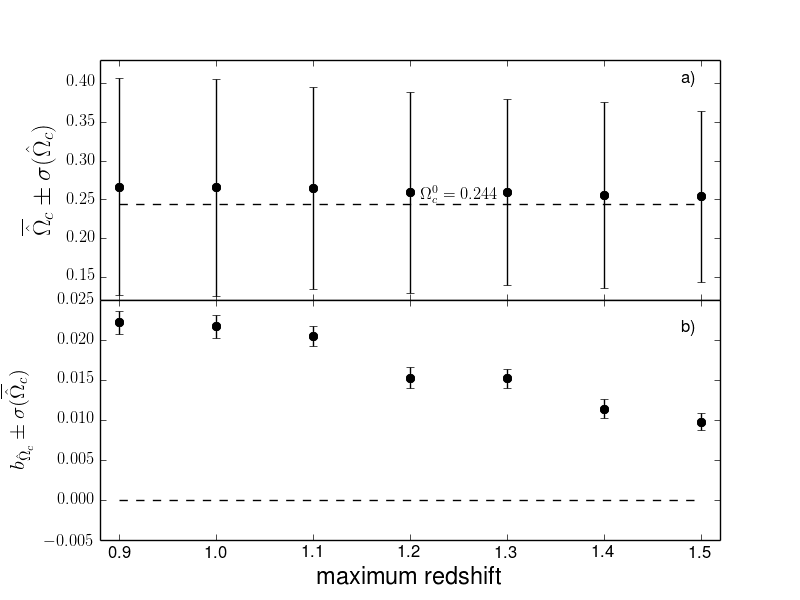}
\includegraphics[scale=0.4]{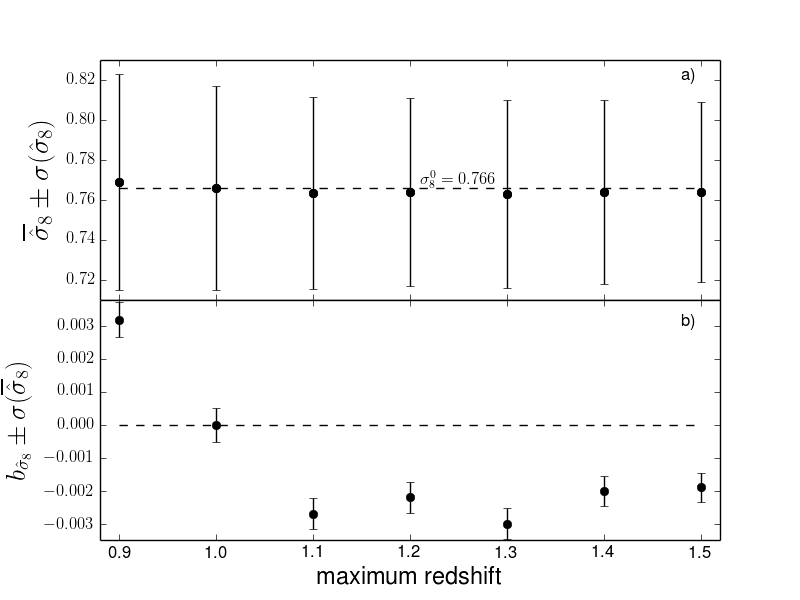}
\begin{center}
\includegraphics[scale=0.4]{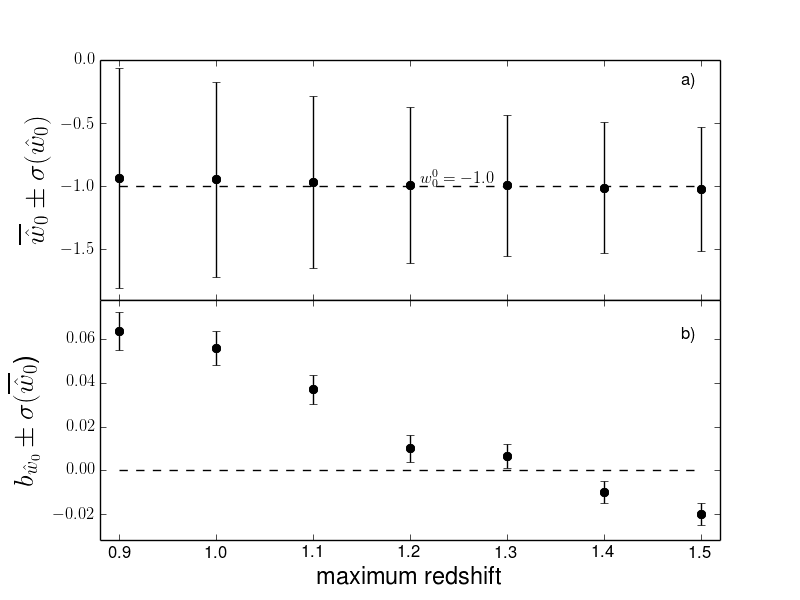}
\end{center}
\caption{The expected values of the estimators (dots) are displayed with the error bars for the estimators (a) and for their means (b) as a function of $z_{max}$.}
\label{fig:redshift_depth}
\end{figure*}

Figure~\ref{fig:redshift_depth} shows the results considering $\Delta\Omega = 720 \, \text{deg}^2$ and the $(\Omega_c, \sigma_8, w_0)$ parametric space. 
The bias on $\hat{\Omega}_c$ decreases systematically with $z_{max}$, whereas $b_{\hat{\sigma}_8}$ decreases and then stays roughly constant after $z = 1.1$. Meanwhile, $b_{\hat{w}_0}$ presents an almost linear behavior as a function of $z_{max}$ with its absolute value increasing after $z=1.3$. In any case, as in the previous plots, the biases are not significant compared to the statistical error bars.

Finally, we study the estimators in the three-dimensional parametric space assuming $z_{max} = 1.5$ and $\Delta\Omega = 10,000$ and $40,000 \, \text{deg}^2$. Both absolute biases and the standard deviations are slightly smaller than those shown in table~\ref{tab:large_areas}, such that $B_{\hat{x}}$ are essentially unaffected by the change in $z_{max}$.  

\subsection{Including photometric redshifts}
\label{sec:bias:photoz}

The analyses carried out on the previous sections assumed perfect knowledge of the redshifts. 
However, spectroscopic redshifts are often available for only a fraction of the clusters, with many of them having $z$ estimated photometrically. 
For example, 49 of 100 clusters from ref.~\cite{Reichardt2013} have just photometric measurements. We therefore include the uncertainties related to photometric redshifts in the likelihood for cluster number counts to verify its effect on the cosmological parameter estimators $\hat{\Omega}_c$, $\hat{\sigma}_8$, and $\hat{w}_0$.

For simplicity we assume that all clusters have the same Gaussian probability distribution for obtaining a photometric redshift $z^{\text{phot}}$ given the true value $z^{\text{true}}$, 
\begin{equation}\label{eq:photo_z}
P(z^{\text{phot}} | z^{\text{true}}) dz^{\text{phot}} = \sqrt{\frac{2}{\pi}} 
 \frac{e^{-\frac{\left(z^{\text{phot}} - z^{\text{true}}\right)^2}{2\sigma_z^2}}}
{\sigma_z (1 - \operatorname{erf} (z^{\text{true}}/\sqrt{2\sigma^2_z}))}  dz^{\text{phot}}, 
\end{equation}
where $\sigma_z = \sigma_z^0 (1 + z)$, $\sigma_z^0 = 0.05$ and the normalization factor is obtained assuming $z^{\text{phot}} > 0$.

Similarly to eqs.~\eqref{eq:ln_eml:unbinned} and \eqref{eq:d2N_z:xi}, we have 
\begin{equation}\label{eq:ln_eml:unbinned:photoz}
\ln \mathcal{L} (\{\theta_j\}, \{\xi_i, z^{\text{phot}}_i\}) = \sum_{i=1}^{n} \ln \left( \frac{d^2N(\xi_i, z_i, \{\theta_j\})}{dz^{\text{phot}} d\xi} \right) - N (\{\theta_j\}) - \ln n!,
\end{equation}
where 
\begin{eqnarray}\label{eq:d2N:photoz:xi}
\frac{d^2N(\xi_i, z^{\text{phot}}_i , \{\theta_j\})}{dz^{\text{phot}} d\xi} &=& \int dz^{\text{true}} \int d\ln M \int d\zeta \, \frac{d^2N(M, z, \{\theta_j\})}{dz d\ln M}   \nonumber \\ 
&\times&  P(z^{\text{phot}}_i | z^{\text{true}}) \, P(\xi_i | \zeta) \, P(\ln\zeta | \ln M),
\end{eqnarray}
and the normalization factor $N (\{\theta_j\})$ is now the expected number of clusters with $z^{\text{phot}}_{min} \leq z^{phot} \leq z^{\text{phot}}_{max}$ and $\xi \geq \xi_{min}$.

Now the MC approach is performed by minimizing eq.~\eqref{eq:ln_eml:unbinned:photoz} with respect to $\theta_j$ and the sampling is carried out using the probability distribution given by eq.~\eqref{eq:d2N:photoz:xi}, the mass-observable relations, and $P(z^{\text{phot}} | z^{\text{true}})$. Each realization is now a catalog of clusters containing their photometric redshifts, $z^{\text{phot}}_i$, and the detection significance, $\xi_i$ (see appendix~\ref{ap:sampling}). 
We study the three-dimensional parametric space for $z_{max} = 1.1$ and $\Delta\Omega = 720 \, \text{deg}^2$.

The results for the relative biases for the 3 parameters are shown in figure~\ref{fig:rel_bias}, together with other cases discussed in this work.
 We find that including the photometric redshift uncertainty has a negligible impact on the relative biases.  
The relative biases are $B_{\hat{\sigma}_c} = 17\%$, $B_{\hat{\Omega}_8} = 6\%$, and $B_{\hat{w}_0} = 8\%$, while the outcomes for the likelihood with only the mass-observable uncertainty give $B_{\hat{\sigma}_c} = 16\%$, $B_{\hat{\Omega}_8} = 6\%$ and $B_{\hat{w}_0} = 5\%$. Thought the biases do increase when considering photometry redshifts, the error bars also increase, keeping the ratio almost constant.
For example, considering photometric redshifts we have $b_{\hat{w}_0} = 0.055$ and $\sigma(\hat{w}_0) = 0.72$, whereas considering perfect knowledge of the redshift yields $b_{\hat{w}_0} = 0.037$ and $\sigma(\hat{w}_0) = 0.68$.

\section{Combination with other observables: CMB distance priors}
\label{sec:cluster:distpriors}

So far we have studied estimators of three cosmological parameters computed using only the cluster abundance. 
However, in practice cosmological constraints are obtained combining cluster counts with other observables, such as the cosmic microwave background (CMB) radiation. Hence, in order to study more realistic estimators, we repeat the MC procedure including CMB data.

We do not consider the full CMB analysis since each case requires 10,000 realizations. Therefore, we build the CMB likelihood using only the so called distance priors (see~\cite{Komatsu2011} and references therein). They are dependent on the cosmological parameters and correspond to the redshift at decoupling $z_\star$, the location of the first acoustic peak $l_A$, and the shift parameter $R$. 

The acoustic scale $l_A$ is given by
\begin{equation}
l_A \equiv (1 + z_{\star}) \frac{\pi D_A(z_\star)}{r_s(z_\star)},
\end{equation}
where $D_A({z_\star})$ is the angular diameter distance and $r_s(z_\star)$ is the sound horizon size.
The shift parameter is
\begin{equation}
R = \frac{\sqrt{\Omega_m H_0^2}}{c} (1 + z_*) D_A(z_*),
\end{equation}
and the decoupling redshift is computed using the Hu \& Sugiyama fitting formula~\cite{Hu1996}.

The likelihood is 
\begin{equation}
\label{eq:CMBL}
-2 \ln L_{CMB} = \sum_{i,j} (x_i - d_i) \left(C_{ij}\right)^{-1} (x_j - d_j),
\end{equation}
where $x_i = (l_A, R, z_\star)$ and $d_i$ are the distance prior values for a given realization (see below). In this work we use \textit{Wilkinson Microwave Anisotropy Probe} 9-year (WMAP9) data, for which the inverse covariance matrix is given in ref.~\cite{Hinshaw2013} as
\[ \left(C_{ij}\right)^{-1} = \left( \begin{array}{ccc}
3.182 & 18.253 & -1.429 \\
18.253 & 11887.879 & -193.808 \\
-1.429 & -193.808 & 4.556 \end{array} \right).\]

We generate a realization for the CMB distance priors $d_i$ assuming that they follow a Gaussian distribution with inverse covariance matrix $\left(C_{ij}\right)^{-1}$ above and $(l_A, R, z_\star)$ calculated from the fiducial model. Therefore, our MC simulations include both the realizations of the cluster distribution and the distance priors, and the likelihood minimized is a combination of eq.~\eqref{eq:CMBL} and the cluster likelihood.

\begin{figure*}
\includegraphics[scale=0.4]{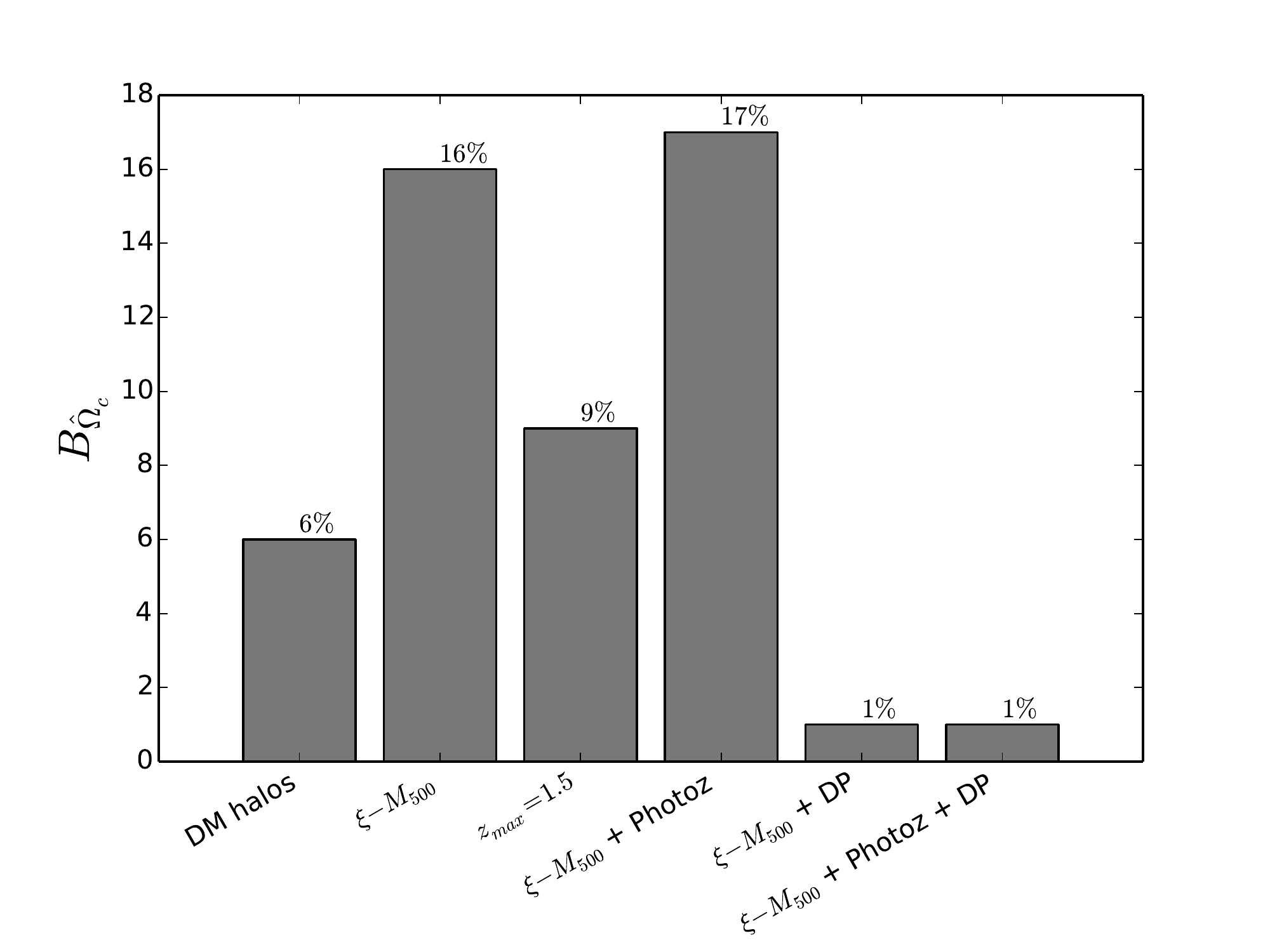}
\includegraphics[scale=0.4]{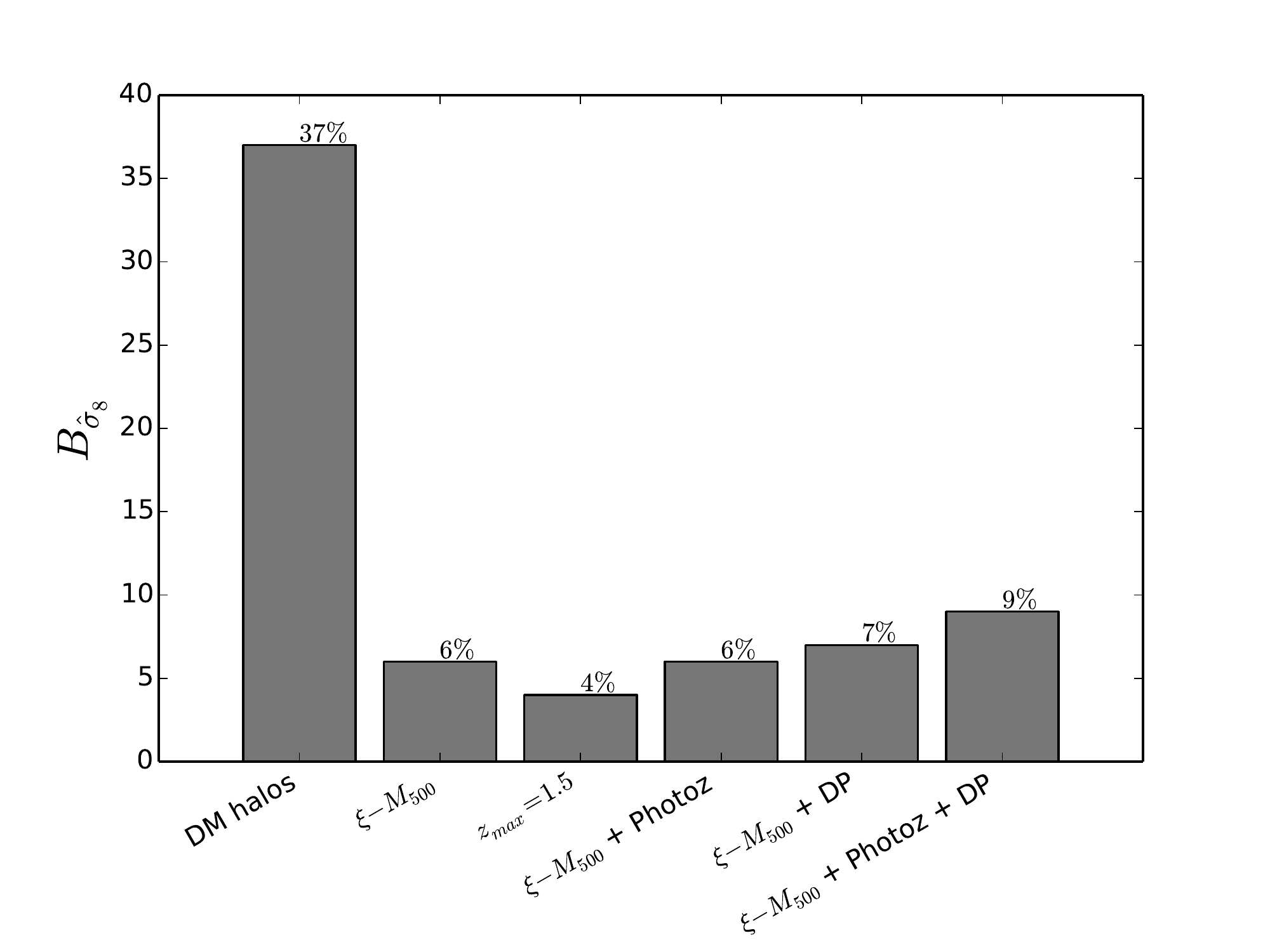}
\begin{center}
\includegraphics[scale=0.4]{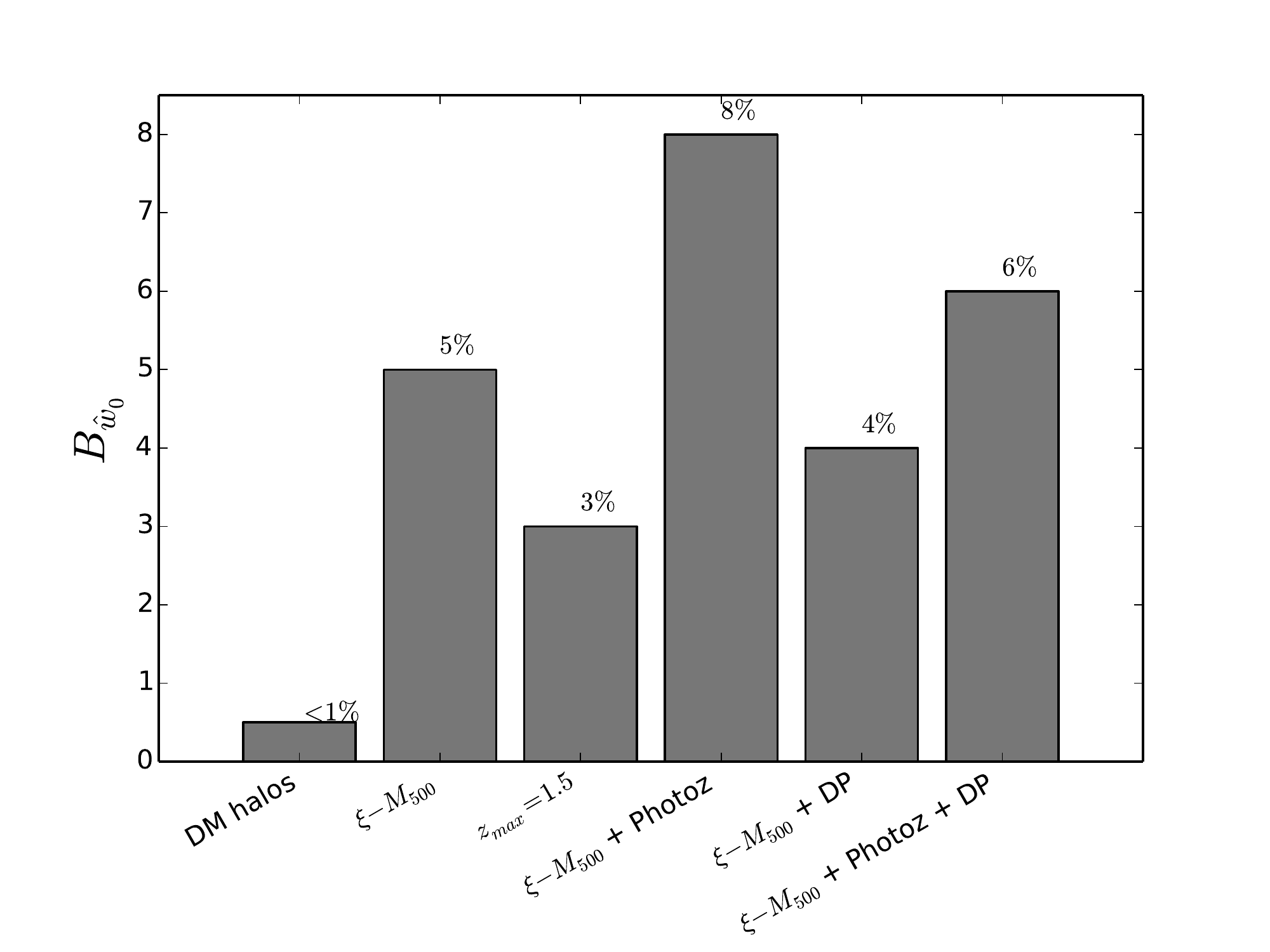}
\end{center}
\caption{The relative biases of $\hat{\Omega}_c$ (left panel), $\hat{\sigma}_8$ (right panel), and $\hat{w}_0$ (lower panel), considering $\Delta\Omega = 720 \, \text{deg}^2$, $z_{max} = 1.1$, and the tridimensional parametric space $(\Omega_c, \sigma_8, w_0)$, obtained with the likelihood (left to right): DM halo abundance, cluster abundance (i.e. including the SZ mass-observable relation), extending the depth to $z_{max} = 1.5$, including photometric redshift uncertainty, combining with WMAP9 distance priors, with and without photometric redshifts, respectively.}
\label{fig:rel_bias}
\end{figure*}

The results from the 10,000 MC realizations for the relative biases $B_{\hat{x}}$ for  $\Delta\Omega = 720 \, \text{deg}^2$, $z_{max} = 1.1$, and the tridimensional parametric space $(\Omega_c, \sigma_8, w_0)$ are shown in figure~\ref{fig:rel_bias}, together with the other cases previously considered in this work. In the combination with the WMAP9 distance priors we consider two cases, including or not the photometric redshifts (always including the SZ mass-observable relation).

The combined likelihood provides a large improvement on both absolute biases and error bars for the $\Omega_c$ and $w_0$ estimators. Roughly, the error bars are 4 to 16 times smaller, depending on the parametric space, than those obtained using only the cluster likelihood.  Assuming $\Delta\Omega = 720 \, \text{deg}^2$ and $z_{max} = 1.1$, we obtain that the biases for $\hat{w}_0$ and $\hat{\Omega}_c$ are small and compatible with zero within their own statistical error bars for the parametric spaces $(w_0)$, $(\sigma_8, w_0)$, and $(\Omega_c, \sigma_8)$. The relative biases also decrease in the $(\Omega_c, w_0)$ plane to $B_{\hat{\Omega}_c} \simeq 10\%$ and $B_{\hat{w}_0} \simeq 5\%$ in comparison to $17\%$ and $9\%$, respectively, obtained with the cluster number counts likelihood only.

The MC results for 178, 720 and $2500 \, \text{deg}^2$ and the three dimensional space $(\Omega_c, \sigma_8, w_0)$ reinforce that $\Omega_c$ is essentially constrained by CMB distance priors, since $\sigma(\hat{\Omega_c}) = 0.01$ and $B_{\hat{\Omega}_c} = 1\%$ for the first two cluster catalogs. We note a stronger weight of the cluster likelihood when $\Delta\Omega = 2500 \, \text{deg}^2$. Besides a decrease in the error bar, there is an increase in the absolute bias such that $B_{\hat{\Omega}_c} \simeq 8\%$. 
In the case of the $w_0$ estimators, the relative biases are $B_{\hat{w}_0} \lesssim 10\%$ for all sky areas, although the biases and error bars are now much smaller.

Differently from $\Omega_c$ and $w_0$, we note from figure~\ref{fig:rel_bias}  that $\sigma_8$ is almost insensitive to the CMB distance priors. The estimators obtained in $(\sigma_8)$, $(\Omega_c, \sigma_8)$, and $(\sigma_8, w_0)$ have the same bias and error bar than that from  unidimensional space using the cluster abundance likelihood, namely $b_{\hat{\sigma}_8} = 0.001$ and $\sigma(\hat{\sigma}_8) = 0.008$, which gives $B_{\hat{\sigma}_8} \simeq 13\%$. In the bidimensional planes, this occurs because $\sigma_8$ is weakly correlated to $\Omega_c$ and $w_0$ ($Cor(\Omega_c, \sigma_8) \simeq -0.25$ and $Cor(\sigma_8, w_0) \simeq 0.09$). On the other hand, in the tridimensional parametric space, the strong correlations between these parameters increase their respective error bars such that $B_{\hat{\sigma}_8} \simeq 10\%$, $7\%$, and $3\%$ for $178$, $720$ and $2500 \, \text{deg}^2$, respectively.

Including photometric redshift uncertainty in the cluster abundance likelihood produces even smaller changes than those verified in section~\ref{sec:bias:photoz}. Assuming $\Delta\Omega = 720 \, \text{deg}^2$, the error bars of the estimators obtained in $(\Omega_c, \sigma_8, w_0)$ space are basically equal to those without considering photometric uncertainty. The biases $b_{\hat{\sigma}_8}$ and $b_{\hat{w}_0}$ are slightly increased such that $B_{\hat{\sigma}_8} = 9\%$ and $B_{\hat{w}_0} = 6\%$, as illustrated in figure~\ref{fig:rel_bias}. 

\section{A comparison of methods to determine confidence contours}
\label{sec:stat_methods}

While the bias on cosmological parameter estimators needs to be evaluated with the Monte Carlo approach, the assessment of statistical errors can also be carried out using other methods such as Fisher matrix (FM) and profile likelihood (PL) (see also~\cite{Holder2001, Wolz2012, Khedekar2013}). Therefore, it is worth comparing the error bars computed with these 3 different methodologies. In ref.~\cite{Khedekar2013}, for example, the authors compare the confidence regions computed via FM and Markov Chain Monte Carlo (MCMC) using only a cluster abundance likelihood, as well with one combined with other observables.

In this section, we briefly review the FM and PL techniques and carry out some examples obtaining error bars and confidence regions for specific realizations.

\subsection{Fisher Matrix}
\label{sec:fisher}

The Fisher matrix is defined as 
\begin{equation}\label{eq:fisher1}
F_{ij} = -\left\langle \left.\frac{\partial^2 \ln \mathcal{L}(\vec{\theta})}{\partial\theta_i\partial\theta_j} 
\right\vert_{\vec{\theta}^0} \right\rangle.   
\end{equation}
If the estimators $\hat{\theta}_i$ are unbiased, then the FM is related with the variance of these estimators as  (see ref.~\cite{Bierens2005})
\begin{equation}\label{eq:fisher2}
C(\hat{\theta}_i, \hat{\theta}_j) = C_{ij} \geq \left[ -\left\langle\left.\frac{\partial^2 \ln 
(L(\vec{\theta}))}{\partial\theta_i\partial\theta_j} \right\vert_{\vec{\theta}^0} \right\rangle \right]^{-1},
\end{equation}
where $C(\hat{\theta}_i, \hat{\theta}_j)$ are the elements of the covariance matrix. In practice it is assumed that the data set is sufficiently large and the second derivative of $\ln \mathcal{L}(\vec{\theta})$ computed at the best-fit $\hat{\theta}$ is a good estimator for the expected value and, therefore, $C_{ij} = (\hat{\mathbf{F}}^{-1})_{ij}$, where $\hat{\mathbf{F}}$ is a $n \times n$ matrix whose elements are $\hat{F}_{ij} = - \left.\frac{\partial^2 \ln (\mathcal{L}(\vec{\theta}))}{\partial\theta_i\partial\theta_j}\right\vert_{\hat{\vec{\theta}}}$. The size of the covariance matrix is defined by the number of parameters to be fitted. Say, for example, that we are fitting $n$ parameters. Thus, the bidimensional contour (ellipse) of the i-th and j-th parameters is computed as~\cite{Coe2009} 
\begin{align}\label{eq:FM_ellipse}
\theta_i(t) = \hat{\theta}_i + q[a\cos(\alpha)\cos(t)- b\sin(\alpha)\sin(t)], \\
\theta_j(t) = \hat{\theta}_j + q[a\sin(\alpha)\cos(t)+ b\cos(\alpha)\sin(t)],
\end{align}
where,
\begin{equation}
a^2 = \frac{C_{ii} + C_{jj}}{2} + \sqrt{\frac{(C_{ii} - C_{jj})^2}{4} + C^2_{ij}},
\end{equation}
\begin{equation}
b^2 = \frac{C_{ii} + C_{jj}}{2} - \sqrt{\frac{(C_{ii} - C_{jj})^2}{4} + C^2_{ij}},
\end{equation}
\begin{equation}
\tan(2\alpha) = \frac{2C_{ij}}{C_{ii} - C_{jj}}.
\end{equation}
The confidence level is defined by $q$, namely, $q = 2.48$ corresponds to $2\sigma$ level. Since $C_{ij}$ is the inverse of an $n\times n$ matrix, the other $n-2$ parameters are marginalized over.
 
This technique is useful to forecast the constraints on cosmological parameters since it has low computational cost. However, the application of the FM  method is limited as it provides just an approximation of the lower bound of the covariance  matrix. 

Although this methodology assumes that estimators are unbiased, we use it to reproduce what is usually done in the literature and then compare the results with MC and PL approaches. It is worth emphasizing that FM can also be obtained for biased estimators~\cite{Ben-Haim2009}. 

We begin the analyses computing the FM of the bi and tridimensional parametric spaces for 20 different realizations, considering the fiducial model with $\Delta\Omega = 720 \, \text{deg}^2$. In general, the FM errors bars are smaller than the those from MC. Only 2, 3, and 5 realizations resulted in larger error bars of $w_0$ in $(\sigma_8, w_0)$, $(\Omega_c, w_0)$ and $(\Omega_c, \sigma_8, w_0)$, respectively. In all these cases, the best-fits were out of the $1\sigma$ confidence interval.

In order to compare the FM with the confidence region computed from MC approach, i.e., from the 10,000 realizations, we choose two realizations where the best-fit of the parameters are within the $1\sigma$ interval in all parametric spaces. The first realization, called \textit{seed 125},\footnote{We use the environment variable GSL\_RNG\_SEED of the GNU Scientific Library (url http://www.gnu.org/software/gsl/) to choose the seed to be used in NumCosmo to randomly generate a catalog.} provides error bars on the cosmological parameters smaller (roughly half) than the MC errors in all studied parametric spaces. The second realization, \textit{seed 131}, gives similar error bars to those obtained with MC approach.
\begin{figure*}
\includegraphics[scale=0.48]{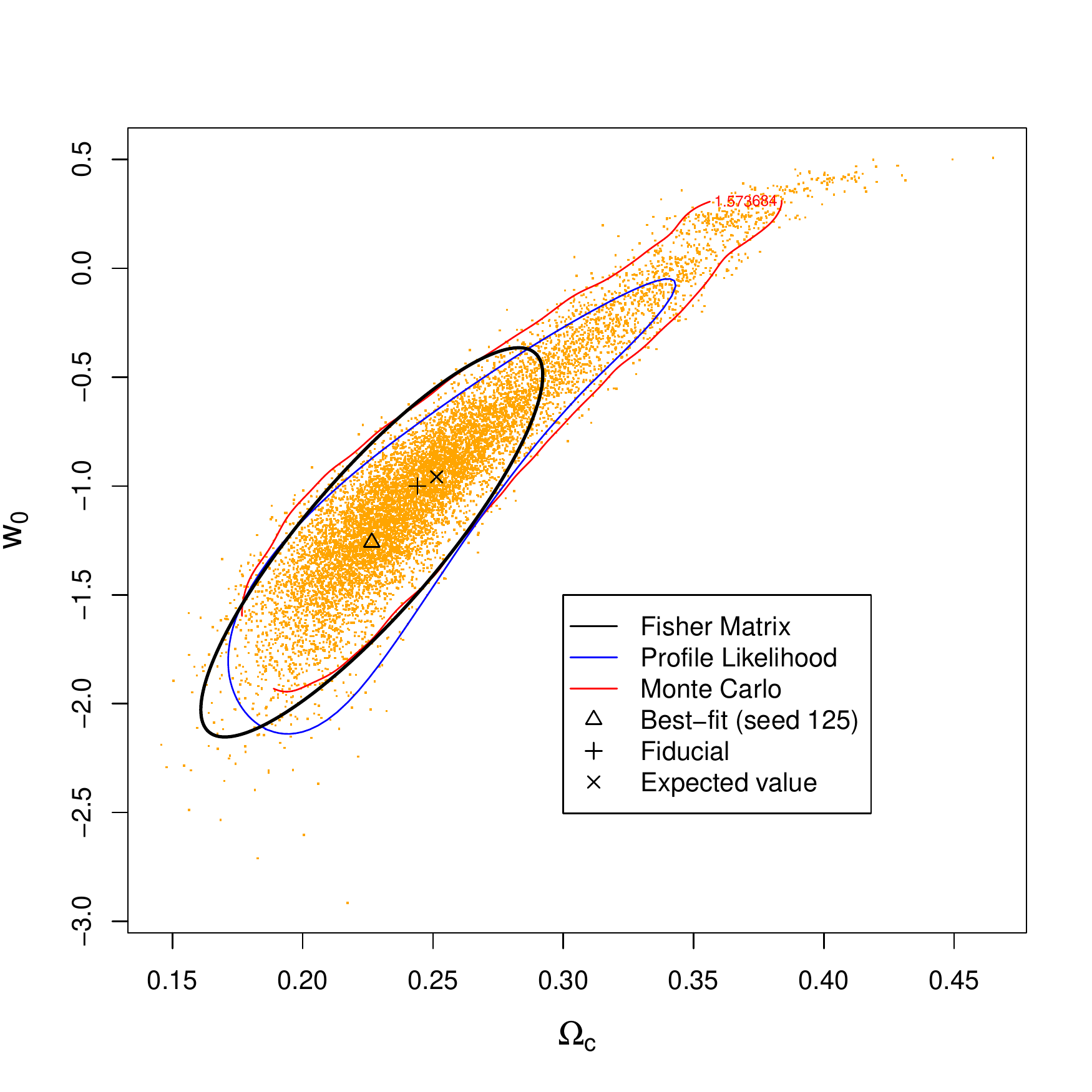}
\includegraphics[scale=0.48]{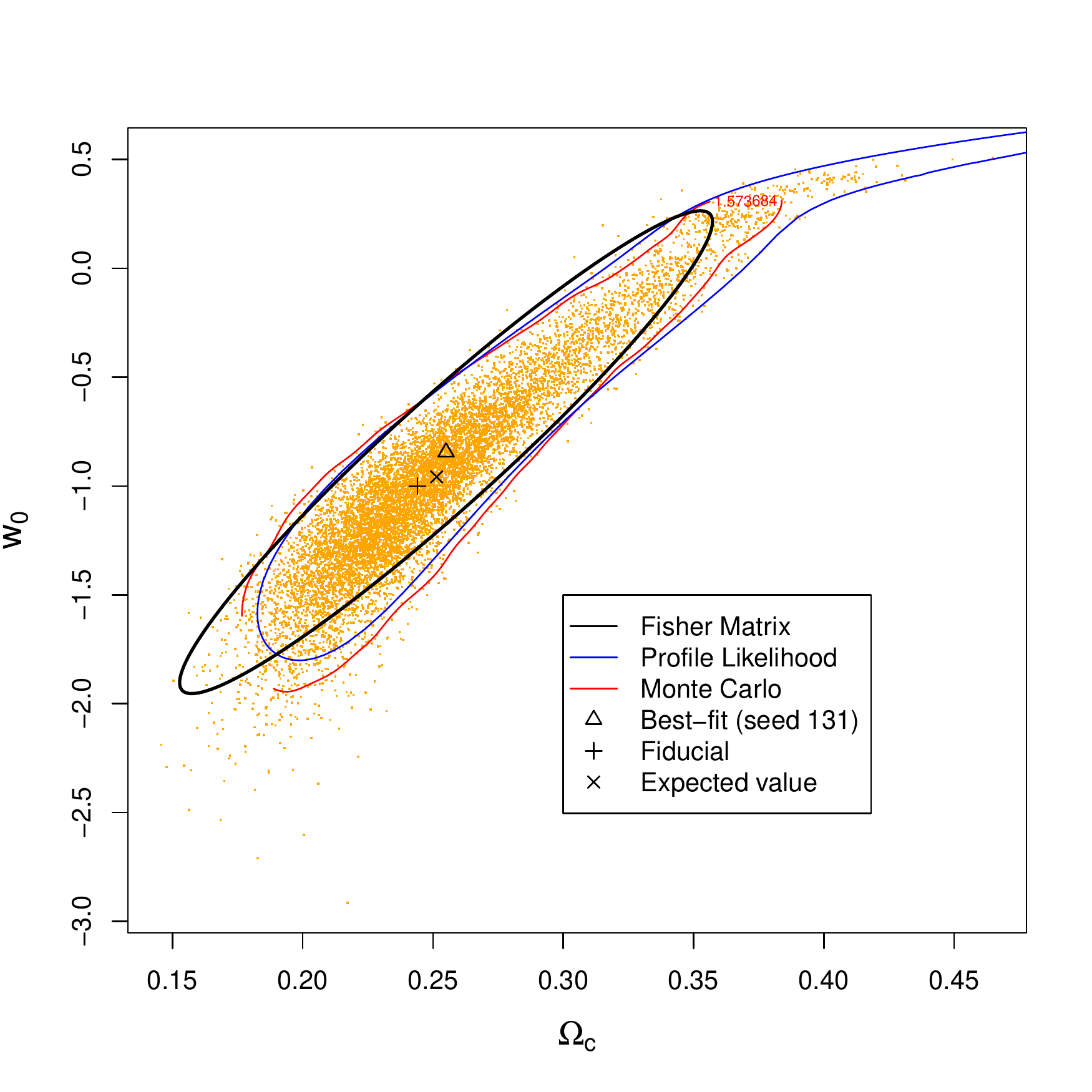}
\caption{$2\sigma$ contour obtained from Monte Carlo approach (red line) and $2\sigma$ confidence regions for two specific realizations (left panel: seed 125 and right panel: seed 131) using Fisher matrix (black line) and profile likelihood (blue line) methods. The orange dots are the best-fits of 10000 realizations.}
\label{fig:contours}
\end{figure*}

Figure~\ref{fig:contours} shows the results for the bidimensional case $(\Omega_c, w_0)$. In both panels, the orange dots represent the best-fits of the 10,000 realizations and the red lines are the $95.45\%$ confidence regions from MC approach. We compute the MC contour using the quantile function from R Project for Statistical Computing.\footnote{http://www.r-project.org/} The black lines are the FM $2\sigma$ bounds for seed 125 (left panel) and seed 131 (right panel) both computed using eq.~\eqref{eq:FM_ellipse}. The FM contours are consistent with the MC result and, as expected, their sizes vary for each realization. 

\subsection{Profile Likelihood}
\label{sec:pl}

The profile likelihood method consists in applying the likelihood ratio test~\cite{Hoel1966, Cowan1998} to obtain the error bounds on the parameters. We define two nested models, i.e., a model with parameters $\{\theta_j\}$ and the same model where the parameters satisfy a set of $n_r$ restrictions $f_a(\{\theta_j\}) = 0$. For each model we obtain the maximum likelihood estimators $\{\hat{\theta}_j\}$ and $\{\hat{\theta}^f_j\}$, where the last are the constrained ones. Note that, in general, these two sets are different. For the first we obtain the maximum of the likelihood with respect to all parameters whereas in the second we obtain the maximum enforcing the conditions $f_a(\{\theta_j\}) = 0$. In a simple example $f(\{\theta_j\}) = \theta_1 - \theta^0_1$, where $\theta^0_1$ is a constant, the restriction is equivalent to keeping a parameter fixed to a constant value. Then, the likelihood ratio test is
\begin{equation}
\label{PLLHR}
\Lambda = \frac{L\left(\{\hat{\theta}^f_j\}, \{y_i\}\right)}{L\left(\{\hat{\theta}_j\}, \{y_i\}\right)},
\end{equation}
where $\{y_i\}$ is the data set. Under certain conditions, $-2\ln\Lambda$ is, asymptotically, a random variable described by a chi-square distribution of $n_r$ degrees of freedom $\chi^2_{n_r}$ (Wilks' theorem)~\cite{Hoel1966, Cowan1998}.

Given the best-fit values $\{\theta^\text{bf}_j\}$, we can use the test described above fixing a given parameter $\theta_k$ to a specific value $\theta^f_k$ to obtain the p-value associated with such choice. 
The p-value is defined as
\begin{equation}\label{eq:p_value}
1-C = \int_{\chi^2_{n_r}}^\infty P_{n_r}(x)dx,
\end{equation}
where $P_{n_r}(x)$ is the probability density of a chi-square distribution of $n_r$ degrees of freedom and $C$ is the confidence interval percentage, e.g., $68.27\%$ (or $1\sigma$ for a Gaussian distribution). Choosing a critical value $C$, we vary the value $\theta^f_j$ to find the interval within which this parameter is accepted for $C$. Thus, the bidimensional contour of $\theta_i$ and $\theta_j$ ($n_r = 2$) is obtained computing eq.~\eqref{eq:p_value}, such that
\begin{equation}\label{eq:pl_2dim}
\chi^2_{n_r=2} = -2\ln\left(\frac{L(\theta^f_i, \theta^f_j, \{\hat{\theta}^f_k\}, \{y_i\})}{L(\{\hat{\theta}^{\text{bf}}_j\}, \{y_i\})}\right),
\end{equation}
where the other parameters $\{\hat{\theta}_k^f\}$ can remain fixed or free. If $\{\theta_k^f\}$ are free, $\{\hat{\theta}^f_k\}$ are the best-fit values computed at $\left(\theta_i^f, \theta_j^f\right)$.

Similarly as done with FM method, we compute some confidence regions using PL. The dashed lines in figure~\ref{fig:contours} correspond to $2\sigma$ contour for seed 125 (left panel) and seed 131 (right panel). 
It is worth noting the ability  of the PL technique in recovering some features, such as the shape, of the real distribution, in contrast to the FM, which, by construction, can only provide an ellipse.
The bottom line is that both FM and PL contours are compatible with the MC result. 

\section{Conclusions}
\label{sec:conclusions}

As has long been anticipated, cosmology has turned from a statistics to a systematics dominated field.
Many sources of systematics have been considered in the literature, in particular those originated from observational effects and the modeling of the physical systems. In this work, we studied the biases from the properties of the estimator itself. Motivated by recent SZ surveys, for which the sample sizes are at most of the order of hundreds of objects and by the fact that ML methods can be biased for small samples (and even for large samples if the estimators are not consistent), we investigated the biases from an unbinned method for cluster number counts, which was shown to be equivalent to the now standard cluster abundance likelihood. The main aim of this work was to determine whether the biases found on some cosmological parameters were significant, as compared to their error bars.
  
To determine the biases from a given likelihood we have used the MC method, generating a large set of simulated samples and comparing the mean values of the cosmological parameters obtained by maximizing the likelihood to the fiducial models used in the simulations.  This is computationally expensive and motivated the development  of optimized codes that were incorporated in the NumCosmo library (see appendix \ref{ap:numcosmo}).
The configurations used in the simulations (survey area, redshift range, cosmological parameters, etc.) were inspired by recent SZ surveys, in particular public catalogs from the SPT.

We first considered an idealized situation where the mass and redshift of the clusters are perfectly known, i.e. the DM halo abundance (section~\ref{sec:no:uncert}). In this case we found that the biases can be quite large for small samples. In particular we obtained a relatively large bias for $\sigma_8$, in the case where $\Delta\Omega = 720 \, \text{deg}^2$ and the likelihood is simultaneously fit for  $\Omega_c$, $\sigma_8$, and $w_0$, as compared to the expected (statistical) error bar on this parameter, yielding $B_{\hat{\sigma}_8} \sim 40\%$. On the other hand, we verified that $\hat{\Omega}_c$, $\hat{\sigma}_8$ and $\hat{w}_0$ are consistent when assuming the full-sky area.

We then considered a mass-observable (SZ) relation from SPT and found that the relative biases are significantly smaller, assuming typical areas of this survey (section~\ref{ParamSpace}). However, the relative biases can be as large as $ \sim 25\%$ of the error bars for some configurations.
Although the absolute values of the bias do in general decrease as the size of the sample increases, at least for the smaller area samples, the statistical error bars also shrink. As a consequence, in these cases, the relative biases are less sensitive to the sample size.

To take a deeper look at the role of sample size, we obtained the biases as a function of both survey area and redshift depth, varying the 3 parameters, $\Omega_c$, $\sigma_8$, and $w_0$  (section~\ref{AreaDepth}).
The smaller absolute biases were obtained for $\Delta\Omega = 1,600$ and $2,500 \, \text{deg}^2$. For larger areas, we found out that the bias grows with the increase of the area. As the standard deviation of the estimators decreases, the relative biases become more relevant. In the full-sky limit, $B_{\hat{x}} \sim 80\%$ and $\sim 50\%$ for one- and three-dimensional parametric spaces, respectively. Therefore, we find that  the estimators computed using $\xi - M_{500}$ cluster abundance likelihood are not consistent, differently from the those obtained with the halo abundance one. In particular, the biases can be a dominant error source as compared to the statistical errors due to shot-noise. However, other sources of error must be considered and it remains to be checked whether the bias found in this work will dominate in this case.

The sensitivity to $z_{\max}$ depends on the parameter under consideration. In any case, for all ranges probed, assuming $\Delta\Omega = 720 \, \text{deg}^2$, the biases are very small compared to the error bars. The situation is practically unchanged when we include photometric redshifts in the MC realizations and the likelihood (section~\ref{sec:bias:photoz}).

Combining the cluster abundance likelihood with CMB distance priors (section~\ref{sec:cluster:distpriors}) evinces that the later imposes strong constraints on $\Omega_c$ and $w_0$, significantly reducing the estimators' error bars in all parametric spaces. The bias is reduced in these cases, specially for $\Omega_c$, since the CMB priors have more weight in this case and are assumed unbiased. 
On the other hand, since $\sigma_8$ is mostly determined by the cluster abundance, 
both its bias and statistical error bar are
roughly unchanged. 

Finally, we have compared the confidence regions of the fitted parameters obtained from the Fisher matrix and profile likelihood methods with the ones derived from the MC realizations and found that the former provide error bars that are broadly consistent with those from MC approach (section~\ref{sec:stat_methods}), further validating the use of these methods in the case of cluster abundance.

It is worth emphasizing that the results obtained are weakly sensitive to the values of the fiducial cosmological model. For example, we used also $\Omega_c = 0.15$, $\sigma_8 = 0.85$, and $w_0 = -0.85$ and found very similar results.

For optical surveys, such as SDSS, DES and LSST, this study must consider sample variance in the cluster likelihood function. Since sample variance dominates the Poisson term for large galaxy clusters samples, we expect that the cosmological parameter estimators would have smaller biases in this context, provided that the sample variance estimator is unbiased.

In general, the mass-observable (M-O) relations are sources of systematic errors in the inference of cosmological parameters. A common approach to minimize the effect of these errors while also constraining the M-O relation is to combine other observables and to fit the cosmological parameters along with those of the M-O relation (so-called self-calibration). Therefore, as future work, we plan to extend this study and obtain the estimators of both type of parameters using, for example, cluster, CMB and type Ia supernovae data. As a preliminary analysis, we used the cluster abundance likelihood and computed the estimators of the two-dimensional parametric spaces, $(\Omega_c, D_{SZ})$, $(\sigma_8, D_{SZ})$ and $(w_0, D_{SZ})$, assuming $720 \, \text{deg}^2$. Following ref.\cite{Benson2013}, we considered a Gaussian prior on $D_{SZ}$ with mean equal to $0.21$ and standard deviation $0.16$. In this case, the cosmological parameters are moderately correlated (anti-correlated) to $D_{SZ}$, $Cor(\hat{x}, D_{SZ}) \simeq \pm 0.5$. Except for $\hat{\Omega}_c$ and $\hat{\sigma}_8$, whose relative biases are about $20\%$, the other estimators have negligible relative biases ($< 5\%$).

If on one hand we obtained that in general $B_{\hat{x}} \lesssim 10\%$ for the most realistic cases of SZ surveys (considering typical areas of SPT), on the other hand we have shown that the consistent feature of the estimators depends on the modeling of the M-O relation and, when this property is not fulfilled, the bias of the estimators cannot be neglected for very large areas. 

The results of this work validate the use of the current maximum likelihood methods for present SZ surveys (small catalogs). However, they highlight the need for further studies for upcoming wide-field sensitive surveys, since biases from SZ estimators do not go away with increasing sample sizes and they may become the dominant source of error for an large area survey, as the Planck probe, and sensitive as the SPT and ACT.

\acknowledgments 

MPL acknowledges CAPES (grant PRODOC 2712/2010) and CNPq (PCI/MCTI/CBPF and PCI/MCTI/INPE programs) for financial support. MM is partially supported by CNPq (grant 309804/2012-4) and FAPERJ (grant E-26/110.516/2012). MPL is grateful to Sandro Vitenti for valuable discussions and comments. The authors thank Fernando de Simoni, Juan Estrada, and Scott Dodelson for useful suggestions. The authors also thank the anonymous referee for the careful reading of this work and valuable suggestions. This work made use of the CHE cluster, managed and funded by ICRA/CBPF, with financial support from FINEP and FAPERJ. 

\appendix 

\section{Numerical Cosmology Library (NumCosmo)}
\label{ap:numcosmo}

In this appendix we present a summary and a short description of each object used in this work. The NumCosmo library is a free library written in the C language dedicated to numerical calculations in cosmology~\cite{Vitenti2012c}. The documentation and examples are available at the project's url.

In section~\ref{sec:counts} we introduced the cosmological model to compute the expected halo number counts. This calculation was split in several objects to allow for each step to be modified for different modeling or optimization. These objects are listed below,
\begin{itemize}
\item NcMatterVar: the filtered matter variance for a given window function [eq.~\eqref{eq:var:R:z}].
\item NcTransferFunc: the transfer function $T(k)$ [eq.~\eqref{eq:powspec}].
\item NcGrowthFunc: the growth function [eq.~\eqref{eq:growth}].
\item NcWindowFunc: the window function [eq.~\eqref{eq:window}].
\item NcMassFunc: the mass function [eq.~\eqref{eq:mass_function}].
\item NcMultiplicity: the multiplicity function [eq.~\eqref{eq:mult}].
\end{itemize}

The NumCosmo's object that implements the sampling method, described in appendix~\ref{ap:sampling}, is NcDataClusterNCount. This object is also used to calculate the abundance likelihood described in eqs.~\eqref{eq:ln_eml}, \eqref{eq:ln_eml:unbinned}, and \eqref{eq:ln_eml:unbinned:photoz}. The mass-observable relation and photometric redshift distribution objects are implementations of NcClusterMass and NcClusterRedshift, respectively. In this work we used the implementations NcClusterMassBenson and NcClusterPhotozGaussGlobal. The WMAP distance priors described in section~\ref{sec:cluster:distpriors} are implemented by the NcDataCMBDistPriors object for both sampling and likelihood evaluation. 

To obtain the results from Fisher Matrix, Monte Carlo, and profile likelihood approaches (Secs.~\ref{sec:fisher} and \ref{sec:pl}, respectively), we used the objects NcmFit for both FM and MC and NcmLHRatio1d for the profile likelihood method.

It is worth emphasizing that, differently from most works, we do not make a grid in redshift and observable quantities, such as $\xi$, to compute eq.~\eqref{eq:d2N_z:xi} [or eq.~\eqref{eq:d2N:photoz:xi}] and, thus, the confidence regions. We calculate these equations at each point $(\xi_i, z_i)$.

\section{Sampling}
\label{ap:sampling}

In this appendix we describe the procedure to generate the realizations of the samples $\{\xi_i, z_i\}$. 
First we compute the halo abundance $N (\{\theta_j\})$ given a fiducial model, 
\begin{equation}
N (\{\theta^0_j\}) = \int_{z_{min}}^{z_{max}} dz \int_{\ln M_{min}}^{\infty} d\ln M \, \frac{d^2 N (\ln M, z)}{dz d\ln M},
\end{equation} 
where $\ln M_{min}$ is obtained from eqs.~\eqref{eq:xi_to_zeta} and \eqref{eq:zeta_to_mass}.
Using $N (\{\theta^0_j\})$  as the mean of a Poisson distribution, we randomly 
generate the total number of objects $n$.

The second step consists in creating a sample with $n$ redshift values $z_i$ from the probability distribution of finding a halo with mass greater than $M_{\text{min}}$ and in the redshift interval $[z, z+dz]$, i.e.,
\begin{equation}\label{eq:pdf_dNdz}
\mathcal{P}(z) dz = \frac{1}{N (\{\theta_j\})}\left(\frac{dN}{dz}\right) dz, 
\end{equation} 
where $N (\{\theta_j\})$ is the normalization factor.
To generate the set $\{z_i\}$ following this distribution we employ the inverse transform sampling, which we briefly describe below. First we define the cumulative 
\begin{equation}\label{eq:zi_uniform}
f(z) = \int_0^z \mathcal{P}(z^\prime) dz^\prime,
\end{equation}
since $f(z)$ is a monotonically increasing function whose image is $[0, 1]$, there is a 
one-to-one relation between $z$ and $f(z)$. Given also that $f(z)$ is a random variable with uniform distribution over $[0, 1]$, we generate $n$ random numbers $\{u_i\}$ from a uniform distribution and inverted equation \eqref{eq:zi_uniform} 
obtaining $z(f)$ and, therefore, the set 
$\{z_i\} = \{z(u_i)\} $.

For each value of $z_i$ generated, we obtain a value $M_i$ from the 
conditional distribution of obtaining a halo with mass in the range $M$ and $M + dM$ given 
$z_i$, defined as
\begin{equation}
P(M|z_i) = \frac{P(M, z_i)}{\mathcal{P}(z_i)},
\end{equation}
where $P(M, z_i)$ is given by eq.~\eqref{eq:pmz_dis} and $\mathcal{P}(z_i)$ by eq.~\eqref{eq:pdf_dNdz}. 
Then we use the cumulative
\begin{equation}
f(M|z_i) = \int_0^{M} P(M^\prime | z_i) dM^\prime
\end{equation}
to define $M(f, z_i)$ and thus we employ the same procedure described above to get $\{z_i\}$, but 
now obtaining $\{M_i, z_i\}$.

Finally, to obtain the cluster sample $\{\xi_i, z_i\}$ (to which we refer as a realization of the cluster 
distribution), a value $\zeta_i$ is randomly generated from the Gaussian distribution $P(\ln \zeta | \ln M)$ with scatter $D_{SZ}$ and mean given by the logarithm base e of eq.~\eqref{eq:zeta_to_mass}, where $M_{500}$ is the true halo mass $M_i$. Then we randomly generate $\xi_i$ using the Gaussian distribution $P(\xi | \zeta)$ with scatter equal to one and mean obtained substituting $\zeta_i$ in eq.~\eqref{eq:xi_to_zeta}. As the last step, we select the objects which fulfill the condition $\xi \geq \xi_{min}$.  

In section~\ref{sec:bias:photoz} we also included the photometric redshift uncertainty. Therefore, the realization  $\{\xi_i, z^{phot}_i\}$ is obtained from the sample $\{\xi_i, z_i\}$ described above where $z^{phot}_i$ is randomly generated from eq.~\eqref{eq:photo_z}. As done for $\xi_i$, the generated values are selected such that $z_{min} \leq z^{phot}_i \leq z_{max}$.


\providecommand{\href}[2]{#2}\begingroup\raggedright\endgroup

\end{document}